\documentclass{emulateapj}
\slugcomment{{\sc Accepted to ApJ:} April 18, 2008}
\def\arcsec{$^{\prime\prime}$}
\newcommand\kms{km s$^{-1}$}

\newcommand\ha{$\rm{H}\alpha$}

\shorttitle{The Kinematics of Thick Disks}
\shortauthors{Yoachim \& Dalcanton}

\begin{document}

\title{The Kinematics of Thick Disks in Nine External Galaxies}

\author{Peter Yoachim\altaffilmark{1,2}
  \&  Julianne J. Dalcanton\altaffilmark{2,3}}
\altaffiltext{1}{Department of Astronomy and McDonald Observatory, University of Texas, Austin, TX 78712; {yoachim@astro.as.utexas.edu}}
\altaffiltext{2}{Department of Astronomy, University of Washington, Box 351580,
Seattle WA, 98195}
\altaffiltext{3}{Tom and Margo Wyckoff Fellow}

\begin{abstract}
We present kinematic measurements of thin and thick disk components in
a sample of nine edge-on galaxies.  We extract stellar and ionized gas
rotation curves at and above the galaxies' midplanes using the Ca {\sc
ii} triplet absorption features and H$\alpha$ emission lines measured
with the GMOS spectrographs on Gemini North and South.  For the higher
mass galaxies in the sample, we fail to detect differences between the
thin and thick disk kinematics.  In the lower mass galaxies, there is
a wide range of thick disk behavior including thick disks with
substantial lag and one counter-rotating thick disk.  We compare our
rotation curves with expectations from thick disk formation models and
conclude that the wide variety of thick disk kinematics favors a
formation scenario where thick disk stars are accreted or formed
during merger events as opposed to models that form thick disks
through gradual thin disk heating.

\end{abstract}

\section{Introduction}

The detailed distribution of stars in galaxies gives vital information
regarding their formation and subsequent evolution.  Of particular
interest are the oldest stellar populations, which in the Milky Way
are the thick disk and halo. These old components provide the best
record of early galaxy assembly.  Originally detected in edge-on S0
galaxies \citep{Burstein79, Tsikoudi79}, thick stellar disks have now
been found in a wide variety of galaxies--S0's \citep{deGrijs96,
deGrijs97b, Pohlen04}, Sb's \citep{Kruit84, Shaw89, vanDokkum94,
Morrison97, Wu02}, and later type galaxies \citep{Dalcanton02,Abe99,
Neeser02, Yoachim06}.  Observations with HST have allowed thick disks
in other galaxies to be studied as resolved population
\citep{Seth05b,Seth07, Tikhonov05, Tikhonovo5b, Mould05}, while
observations at high redshift show potential thick disks in the
process of forming \citep{Elmegreen06}.

The most detailed studies of thick disks come from observations within
the Milky Way.  Since its discovery \citep{Gilmore83}, the MW thick
disk has been found to be structurally, chemically, and kinematically
distinct from the thin disk.  Structurally, star counts with large
surveys such as SDSS and 2MASS reveal the galaxy is best fit with two
disk components \citep[e.g.,][]{Ojha01,Juric08}.  Chemically, thick
disk stars are more metal-poor and older than stars in the thin disk
\citep[e.g.,][]{Reid93, Chiba00}.  They are also significantly
enhanced in $\alpha$-elements, compared to thin disk stars of
comparable iron abundance \citep{Prochaska00, Taut01, Bensby03,
Feltzing03, Mishenina04, Brewer04,Bensby05,Brewer06,Ram07}.
Kinematically, thick disk stars have both a larger velocity dispersion
and slower net rotation than stars in the thin disk \citep{Nissen95,
Chiba00, Gilmore02, Soubiran03, Parker04, Girard06}.  All of these
facts lead to the conclusion that the thick disk is a relic of the
young Galaxy.  As such, it provides an excellent probe of models of
disk galaxy formation (see reviews by \citet{Nissen03, Freeman02}).

Given these systematic differences between their properties, thick and
thin disks are likely to have distinct formation mechanisms.  The
structure, dynamics, and chemical abundance of the thin disk strongly
suggest that the majority of its stars formed gradually from a thin
rotating disk of high angular momentum gas \citep{Fall80,
Chiappini97,Cescutti07}.  In contrast, the formation of the thick disk
is still poorly constrained and is likely to be more complex.

Thick disk formation models can be grouped into three broad
categories.  In the first, a previously thin disk is kinematic heated.
In this scenario, stars form in a thin disk and increase their
velocity dispersion with time.  This vertical heating can be rapid,
due to interactions and mergers \citep{Quinn93, Walker96, Velazquez99,
Chen01,Robin96} or gradual, due to scattering off giant molecular
clouds, spiral arms, and/or dark matter substructure
\citep{Villumsen85, Carlberg87, Hann02, Benson04, Hayashi06, Kaz07}.
In the second formation scenario, stars ``form thick'' with star
formation occurring above the midplane of the galaxy \citep{Brook04}
or form with large initial velocity dispersions in large stellar
clusters \citep{Kroupa02}.  In the final class of models, thick disk
stars are directly accreted from satellite galaxies.  Numerical
simulations have shown that stars in disrupted satellite galaxies can
be deposited onto thick disk like orbits \citep{Abadi203, Martin04,
Bekki01, Gilmore02, Navarro04,Statler88}, producing extended stellar
debris such as seen around M31 \citep{Ibata05,Kalirai06,Ferguson02}.
While these models were originally developed to explain the origin of
the MW thick disk, they should work equally well for thick disks in
other galaxies.

Measuring the kinematics of thick disk stars is one of the best
discriminators between the formation models.  If the thick disk forms
from a heated thin disk, we expect the kinematics of the two
components to be closely related.  On the other hand, if the thick
disk stars form outside the galaxy and are later accreted, we could
find systems where the thick disk kinematics are completely decoupled
from the thin disk.

In this paper, we present observations of stellar and gas kinematics
in nine edge-on systems as part of our continuing analysis of thick
disks in a large sample of edge-on galaxies \citep{Dalcanton00}.
Compared to \citet{Yoachim05}, which presented the first two galaxies
in this study, we have improved the analysis techniques and
significantly expanded our sample size.

\section{Observations}

\subsection{Target Selection}
We have carried out long-slit spectroscopic observations using the
Gemini North and South telescopes of nine galaxies drawn from the
\citet{Dalcanton00} sample of edge-on late-type galaxies.  The
original sample of 49 galaxies was selected from the Flat Galaxy
Catalog \citep{Karachentsev93} and imaged in $B$, $R$, and $K_s$
\citep{Dalcanton00}.  This sample was selected to contain undisturbed
pure disk systems spanning a large range of mass.  \citet{Dalcanton02}
used this imaging to demonstrate the ubiquity of thick disks around
late-type galaxies, while \citet{Yoachim06} used two-dimensional
photometric decompositions to measure the structural parameters for
the thick and thin disks.  All the galaxies in the sample presented
here have prominent thin star forming disks.

Our spectroscopic program targeted galaxies spanning a wide range of
masses ($50<V_c <150$ \kms).  The sample targets were limited to those
that had thick disks that we believed we could isolate
adequately--i.e., those that had significantly larger scale heights
from the thin disk and that were bright enough that we could acquire
spectra in reasonable observing times.  This constraint caused several
of the higher mass galaxies to be rejected from the kinematic sample,
as the regions where the thick disk could be expected to dominate were
simply too faint.  This bias is consistent with the conclusion of
\citet{Yoachim06} that the thick disk is more prevalent in lower mass
galaxies.  Our selection criterion limited the sample to $\sim$20
galaxies of the original 49.  We also selected galaxies to be at
redshifts such that the Ca features did not land on night sky emission
lines.  In our initial observations, we submitted more galaxies than
we could observe and let the Gemini observing specialists select which
galaxies would best fit with the queue scheduling.  For the final
observing runs we explicitly selected galaxies to ensure that a
reasonable mass range was observed in the final sample.  The
properties of the final sample are listed in Table~\ref{gal_list}.

\subsection{Observing Strategy}

Based on the thin and thick disk decompositions in \citet{Yoachim06},
we targeted regions of the galaxies where the flux is dominated by
either the thin or thick disk stars.  The two highest mass galaxies in
our sample have notable dustlanes \citep{Dalcanton04}, and for these
we offset the spectra slit to observe regions of the galaxy which
should be optically transparent.  We discuss possible residual dust
effects in detail in \S~\ref{Sdust}.  When selecting slit placement
for the offplane, the direction of offset was based primarily on
avoiding foreground objects and the ability to use a single guide star
for all dither positions.

For our instrumental setup, we used GMOS on Gemini North in longslit
mode with a 0.5\arcsec\ slit and the R400\_G5305 grating set to a
central wavelength of $\sim8440$ \AA\ along with the OG515\_G0306
filter.  Similarly for observations from Gemini South, we used a
0.5\arcsec\ slit the R400+\_G5325 grating and OG515\_G0330 filter.
For both GMOS setups, we binned the CCDs by 2 in the spatial direction
during readout giving a pixel scale of 0.145\arcsec/pix in the spatial
direction and 0.69 \AA/pixel in the spectral direction.  The resulting
spectra cover the wavelength range of $\sim 6330-10570$ \AA, although
there is heavy residual fringing redward of 9300 \AA.  Exposure times
for individual frames were 900, 1200, or 1800 seconds.  The midplanes
were observed 3-5 times while offplane positions were observed 18-51
times depending on the galaxy.  Exposures were spatially dithered
$\sim$30\arcsec\ along the slit.  These configurations allow us to
simultaneously observe the H$\alpha$ emission and Ca {\sc ii} triplet
absorption features out to large radii.

All of the observations were executed in queue mode over five
semesters.  The observation details for each galaxy are listed in
Table~\ref{obstable}, with details of the slit positions listed in
Table~\ref{slittable}.

%\clearpage
\begin{deluxetable*}{ c c c c c c c}
\tabletypesize{\scriptsize}
\tablecaption{Properties of Targeted Galaxies \label{gal_list} }
\tablewidth{0pt}
\tablehead{
\colhead{Galaxy} & \colhead{Dist$^1$} & \colhead{V$_c$} & \colhead{$h_R$} & 
\colhead{$z_{0,thin}$} & \colhead{$z_{0,thick}$} & \colhead{$L_{thick}/L_{thin}$}
\\
\colhead{FGC} & \colhead{Mpc} &\colhead{\kms} & \colhead{\arcsec}& 
\colhead{\arcsec}& \colhead{\arcsec}
}
\startdata
227 &  89.4 &  106.0 &   10.2 &    1.8 &    3.9 &   0.47  \\
780 &  34.4 &   75.0 &   15.1 &    3.1 &    8.4 &   0.93  \\
1415 &  38.3 &   86.5 &   18.3 &    2.8 &    6.6 &   0.95  \\
1440 &  70.9 &  150.5 &   15.9 &    2.3 &    5.0 &   0.38  \\
1642 &  36.6 &   55.0 &   12.5 &    3.1 &   10.0 &   0.19  \\
1948 &  36.9 &   54.5 &   12.3 &    1.6 &    3.6 &   3.56  \\
2558 &  73.8 &   89.0 &    9.2 &    2.6 &    3.6 &   0.47  \\
E1371 &  82.6 &  131.0 &    7.7 &    1.6 &    3.4 &   0.27  \\
E1498 & 135.5 &  133.0 &    7.6 &    1.2 &    3.8 &   0.19  \\
\enddata \\
%\begin{flushleft}
{$^1$\citet{Kara00} }
%\end{flushleft}

\end{deluxetable*}

%--------------------------------------
\begin{deluxetable*}{lllcc}
\tabletypesize{\scriptsize}
\tablecaption{Observing Details \label{obstable} }
\tablewidth{0pt}
\tablehead{
\colhead{Galaxy} & \colhead{Gemini ID} & \colhead{Observation Dates} & 
\colhead{Midplane Exposure} & \colhead{Offplane Exposure} \\
\colhead{FGC}  &   & & \colhead{\# x time (s)}& \colhead{\# x time (s)}} 
\startdata
 1415 & GN-2003A-Q-6 & 3-28-2003 to 06-06-2003 & 3x900 & 41x1200 \\%& 0 & 5.4, 1.0 \\
 227 &GN-2003B-Q-51 &  9-21-2003 to 11-22-2003 &   3x1200 &  27x1200 \\%& 0 & 3.0, 1.3\\
 1642 &GN-2004A-Q-54 & 02-16-2004 to 06-24-2004 & 3x1200 &  51x1200 \\%& 0 & 4.2, 0.75 \\
 780 &GN-2004A-Q-54 & 02-20-2004 to 04-27-2004 & 5x1200 &31x1200 \\%& 0 & 6.5, 1.1\\
 2558 &GN-2004B-Q-29& 07-15-2004 to 11-20-2004 & 3x1200 & 36x1200 \\%& 0 & 3.9,1.4\\
 E1498&GS-2004B-Q-44& 03-11-2005 to 06-10-2005 & 3x1200 & 50x1200 \\%& 0 &2.0, 1.3\\
 1948 &GN-2005A-Q-21& 08-12-2004 to 08-24-2004 & 5x1800 & 18x1800 \\%& 0 & 3.1, 0.55\\
 E1371&GS-2005A-Q-17& 04-05-2005 to 04-14-2005 & 3x1200 & 21x1800 \\%& 0.5 &0.2& 2.8,1.1\\
 1440&GS-2005A-Q-17& 02-11-2005 to 04-05-2005 & 3x1200 & 30x1800 \\%& 0.5 & 0.2& 4.5,1.5\\
\enddata \\
%\begin{flushleft}
%{\scriptsize any other little note here}
%\end{flushleft}
\end{deluxetable*}

%--------------------------------------
\begin{deluxetable*}{lllllllll}
\tabletypesize{\scriptsize}
\tablecaption{Slit Placement \label{slittable} }
%\tablewidth{0pt}
\tablehead{
\colhead{Galaxy} &  \multicolumn{2}{c}{Midplane Offset$^1$} 
&\multicolumn{2}{c}{Offplane Offset} \\
 \colhead{FGC} & \colhead{ arcsec}&\colhead{ kpc} & \colhead{ arcsec}
& \colhead{ kpc} 
& \colhead{$z/z_{0,thin}$} & \colhead{$z/z_{0,thick}$} }
\startdata
227 &   0.0 &   0.0 &   3.0 &   1.3 &   1.7 &   0.8  \\
780 &   0.0 &   0.0 &   6.5 &   1.1 &   2.1 &   0.8  \\
1415 &   0.0 &   0.0 &   5.4 &   1.0 &   1.9 &   0.8  \\
1440 &   0.5 &   0.2 &   4.5 &   1.5 &   2.0 &   0.9  \\
1642 &   0.0 &   0.0 &   4.2 &   0.7 &   1.4 &   0.4  \\
1948 &   0.0 &   0.0 &   3.1 &   0.6 &   1.9 &   0.9  \\
2558 &   0.0 &   0.0 &   3.9 &   1.4 &   1.5 &   1.1  \\
E1371 &   0.5 &   0.2 &   2.8 &   1.1 &   1.8 &   0.8  \\
E1498 &   0.0 &   0.0 &   2.0 &   1.3 &   1.7 &   0.5  \\
\enddata \\
%\begin{flushleft}
{$^{1}$Midplane offset to avoid obvious dust lanes.}
%\end{flushleft}
\end{deluxetable*}
%\clearpage

\subsection{Data Reduction}

A combination of Gemini IRAF packages, standard IRAF packages, and
custom IDL code were used to reduce our data.  These procedures have
been improved since initial results for FGC 227 and FGC 1415 were
published in \citet{Yoachim05} and have been applied to the entire
data set.  We bias corrected the images using a fit determined from
the overscan region followed by subtracting residual structure
measured from a bias frame.  Because both GMOS North and South are
extremely stable, we were able to create average bias images by
combining $\sim$60 bias frames per observing semester.  We
interpolated the three GMOS chips into a single image using the Gemini
IRAF tasks, after which the standard IRAF reduction tools were used.
For Gemini-South observations, we also needed to subtract a dark
current correction of $\sim$6-12 counts from the science frames.
Gemini-North images showed no detectable dark current.  Images were
flat-fielded using GCAL lamp flats that were taken every hour
interspersed with the science observations, minimizing the amount of
fringing present in the final frames.  We applied a slit illumination
correction using twilight sky observations.

For wavelength calibration, we used the night-sky atlases of
\citet{Oster96} and \citet{Oster97} to create a sky line list
containing only lines (or stable unresolved doublets) that could be 
centroided with our instrumental set-up.  For each science exposure,
we identified 100-110 sky lines to use for rectification.  We then used
these lines for a 5th order Legendre polynomial fit for wavelength
calibration, and rebinned our spectra to a common dispersion.  Typical
dispersions were 0.69 \AA\ pixel$^{-1}$ with calibration arc lamps showing a
FWHM of 3.8 \AA.  The wavelength solutions were stable over each
observing night.

Sky subtraction proved difficult because of the large number of strong
sky emission lines.  If we use standard sky subtraction techniques, we
find that there are large systematic residuals left on our frames due
to variation in the width of the slit along its length. 
The RMS deviation in the centroid position of a single sky line is
$\sim$0.07 \AA\ while the RMS of its Gaussian FWHM is 0.11 \AA.  This
is a surprisingly high variation for the width of the slit.  We have
tried the sky-subtraction techniques described in \citet{Kelson00} and
find that the systematic residuals remain, although the
\citet{Kelson00} sky-subtraction technique does eliminate problems
associated with wavelength rectification and interpolation. Having
eliminated our data-reduction procedure as the cause, we conclude the
high dispersion in sky line FWHM is indicative of a systematically
varying slit width.  In many cases, such residuals can be removed
using the nod-and-shuffle technique \citep{Glazebrook01}.
Unfortunately, our galaxies are too large ($\sim$1 arc minute, or 1/3
of the total slit width) to make effective use of traditional
nod-and-shuffle.

\begin{figure*}
\epsscale{1}
\plottwo{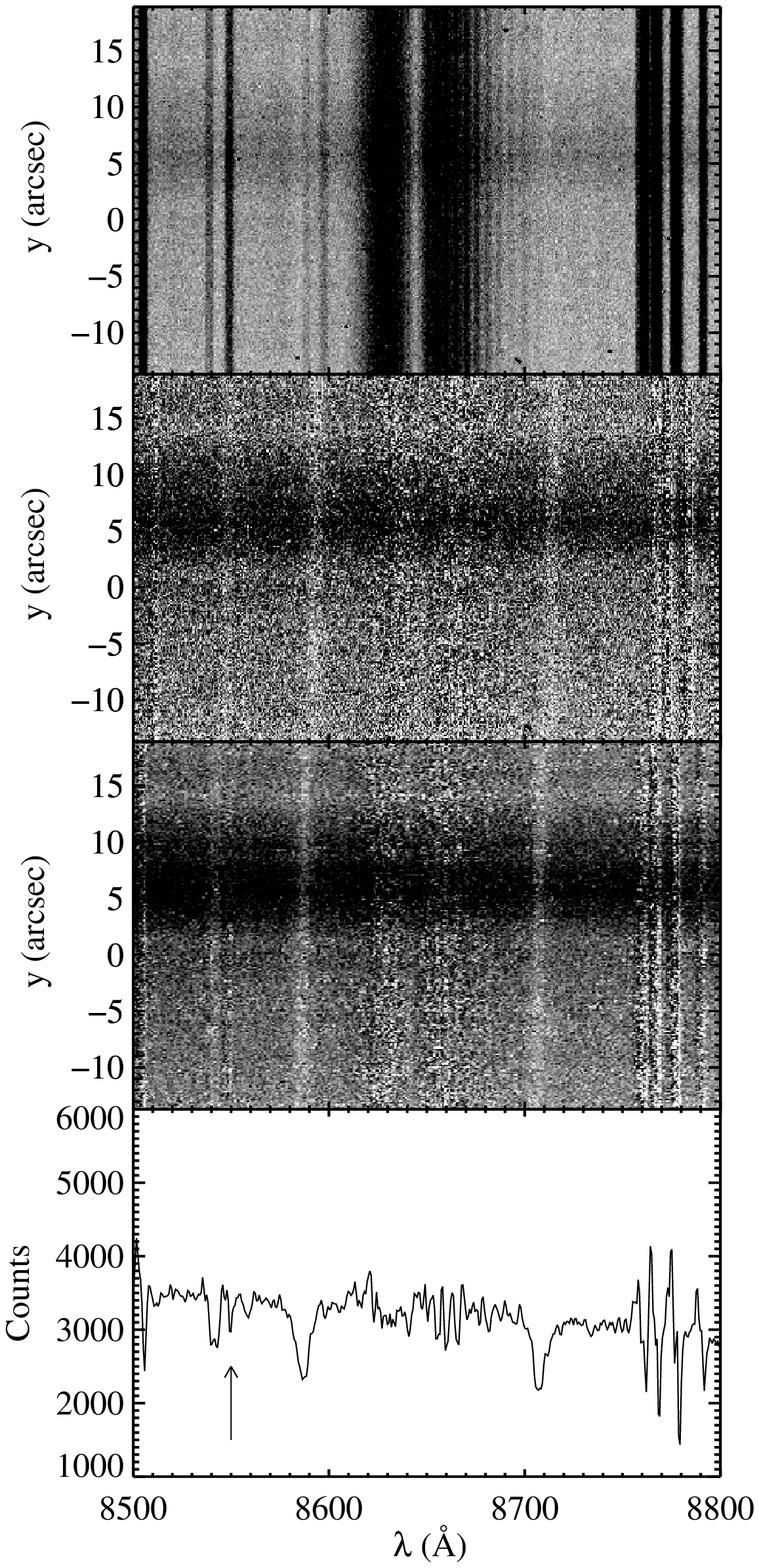}{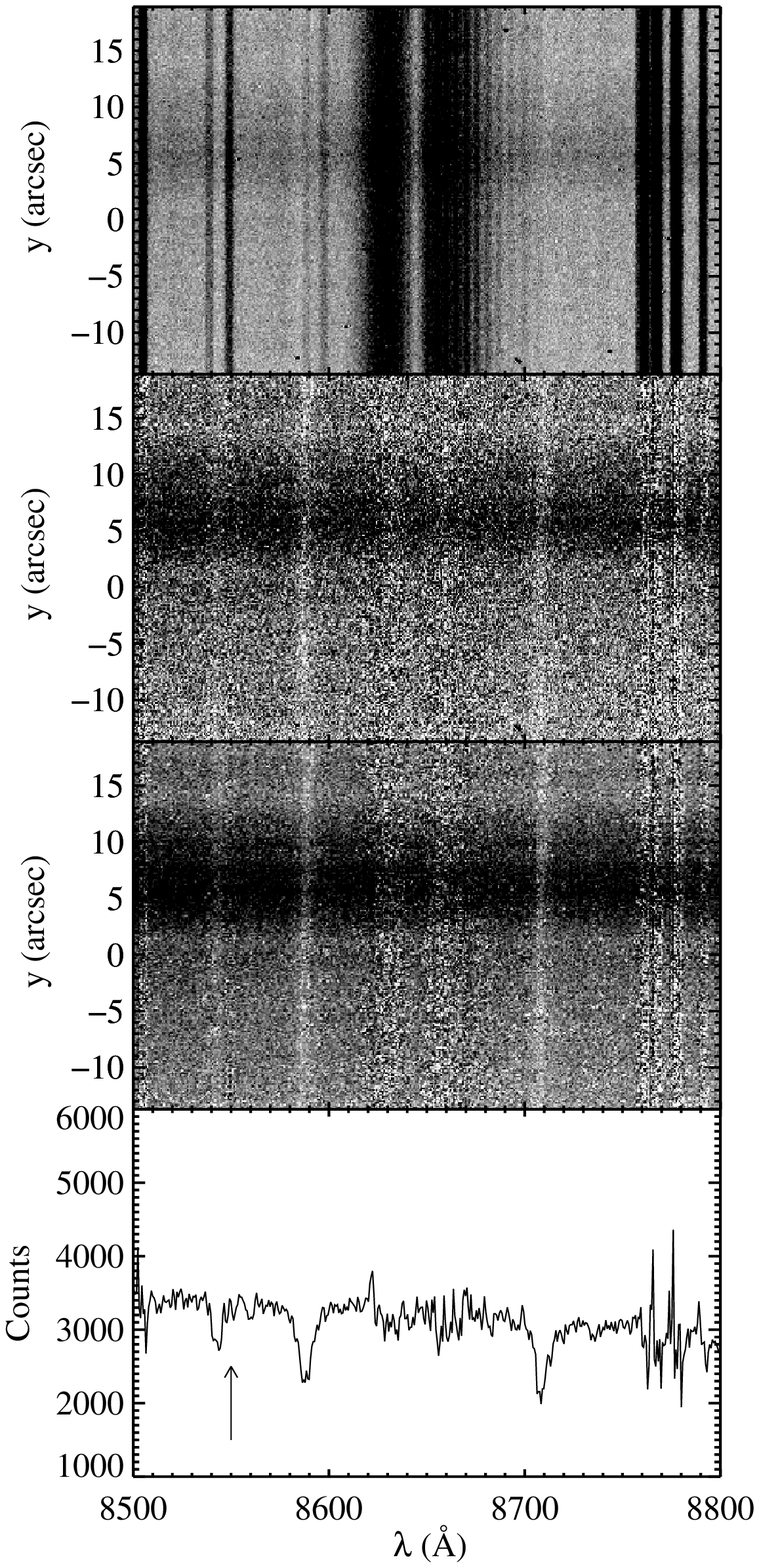}
\caption{Results from different sky subtraction techniques for the
midplane of FGC 1415.  On the left, we show the results from standard
sky subtraction techniques and the right panels show our sky template
subtraction.  Top panels show the raw galaxy spectrum before the sky
has been subtracted.  Middle panels show a single subtracted frame and
the final combined image.  The bottom panel shows the combined
spectrum summed along the spatial dimension.  An arrow points out a
sky line residual present in the standard subtraction that is
eliminated in template subtraction.  The brightest sky lines leave
large residuals in both cases, but the magnitude of residuals is
decreased significantly with the nod-and-shuffle-like technique (see
the lines near $\sim8770$ \AA, for example).  \label{sky_sub} }
\end{figure*}

To remove the systematic residuals present in the bright sky lines, we
employ a nod-and-shuffle like template subtraction.  Because we placed
different galaxies on different spatial sections of the chips, all of
the slit was illuminated by sky for at least some observations.  We
therefore could construct high S/N sky frames by masking objects in
our 2-d spectra and combining the wavelength rectified frames.  By
doing this, we create a deep sky frame for each observing quarter.  We
then remove the sky background by selecting a sky-dominated region in
a science frame and scaling the sky image column-by-column to match
the science frame sky region, then subtract the rescaled sky frame
from the science image.  In most cases, we were forced to apply sky
frames generated from different observing semesters to the science
frames.  Luckily, our instrument setup quarter-to-quarter was
identical, and the GMOS instruments are stable enough that this
technique works well at removing systematics caused by the variable
slit width.  This sky subtraction technique appears to give results
comparable to nod-and-shuffle technique for individual frames.  Our
sky subtraction procedure incurs a small signal-to-noise penalty, but
is effective at removing the systematic residuals from moderate
sky lines (Figure~\ref{sky_sub}).

\begin{figure*}
\plotone{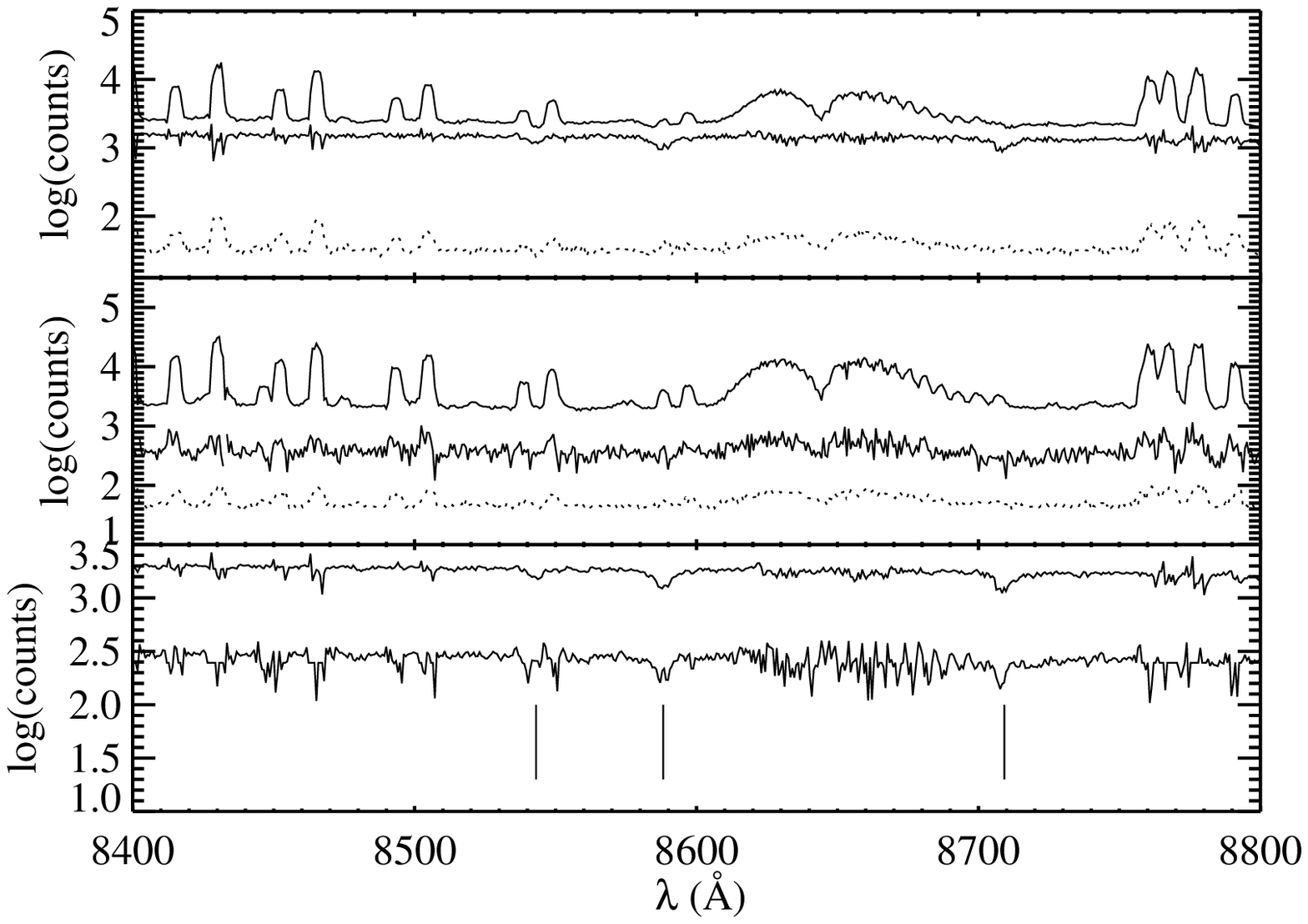}
\caption{Examples of spectra before and after sky subtraction.  The
top panel shows the results of a single midplane exposure before and
after sky template subtraction (top and middle curves respectively).
The middle panel shows a single offplane exposure before and after
extraction.  Dotted lines show the RMS noise level in the spectra.
The bottom panel shows the final midplane and offplane spectra after
all the frames have been averaged together.  The largest systematic
residuals from the sky lines have been masked.  The three vertical
marks show the location of the Ca triplet absorption lines.  All of
the spectra were extracted over the central 14\arcsec\ of the galaxy.
\label{sky_level} }
\end{figure*}

This excessive agonizing over sky subtraction is demanded by the very
low surface brightness levels of our targets.  For an individual
midplane image, the brightest part of the galaxy is $\sim20\%$
brighter than the sky level, and for individual offplane images the
signal is only $\sim11\%$ the sky background.  Examples of the spectra
extracted over the central 14\arcsec\ spatial extent of the galaxy
before and after sky subtraction are shown in Figure~\ref{sky_level}.

When we combine several hours of observations we are more sensitive to
low surface brightness features, and find some wavelengths are still
dominated by systematic noise.  Even with our sky template correction,
some sky lines are so bright that some systematic residuals remain.
When we use conventional sky-subtraction techniques, residual errors
have maximum deviations of $\pm$55\% while the template subtraction
gives deviations of $\pm$38\%.  While deviations of 38\% swamp out the
signal from any stellar absorption lines near bright sky lines, the
residual deviations for smaller sky lines are decreased to a level
where the stellar absorption lines can be accurately measured.  In
Figure~\ref{sky_sub}, we compare the two sky subtraction routines.
The extracted spectra look similar, with both being dominated by the
sky line residuals redward of 8750 \AA.  The template subtraction is
able to eliminate the residuals left from the sky line at 8555 \AA,
just to the right of the weakest Ca {\sc ii} triplet line, and reduces
the large residuals at the reddest wavelengths plotted.

%\clearpage

%\clearpage

After the sky had been removed, the images were Doppler-corrected for
motion relative to the Local Standard of Rest and combined.  Before
cross-correlation was performed, the spectra were rebinned into
logarithmic wavelength bins.

\section{Rotation Curves}

\subsection{H$\alpha$ Rotation Curves}

Both our midplane and offplane observations show strong H$\alpha$
emission.  For each galaxy, we extracted a series of 1-D spectra by
summing 28 pixels ($\sim 4$\arcsec) along the spatial dimension.  The
ionized gas rotation curve was fit with a Gaussian peak to the
H$\alpha$ line.  In principle, an envelope-tracing method would
produce a more robust measure of the rotation curve.  However, we find
that the width of the H$\alpha$ lines (FWHM$\sim$3.8 \AA) are identical
to the instrumental dispersion as measured from the arc lamps
(FWHM$\sim$3.8 \AA), and we would thus not gain much accuracy from a
more detailed rotation curve extraction.

The [N{\sc ii}] and [S{\sc ii}] lines are present as well, but the
H$\alpha$ line is so strong that we found no additional advantage in
fitting all the emission lines simultaneously.  We find typical
uncertainties in the central wavelength of the H$\alpha$ Gaussian peak
of 1-2 \kms ~for midplane observations and 4-7 \kms ~for offplane
observations.

To double check the accuracy of our extracted rotation curve, we fit
rotation curves to night sky lines before the background is subtracted
off.  Perfect calibration would result in sky line rotation curves
with zero rotation.  The central wavelengths of the sky lines vary
with an RMS error of 2.4-3.5 \kms, with the higher value resulting
from larger spatial extraction windows.  Most of this scatter can be
attributed to uncertainties in the wavelength rectification solution.
With fewer sky lines around H$\alpha$ compared to the redder regions
of our spectra, the rectification is not as well constrained.
Overall, these tests suggest that we are able to extract the ionized
gas rotation curve with an error of a few \kms.

The resulting H$\alpha$ rotation curves are plotted as solid lines in
Figure~\ref{all_rc}.  Our data show a tight agreement between the
midplane and offplane H$\alpha$ curves, which is a good sign that dust
is not obscuring the midplane rotation curves.  If we were observing
along major dustlanes, we could expect to see the offplane
observations rotating faster than the midplane, especially at small
galactic radii (see \S~\ref{Sdust}).

We leave a detailed analysis of the gas kinematics for a later paper.
At this time, we simply note that the midplane and offplane H$\alpha$
rotation curves are surprisingly well matched.  This is slightly
unexpected, as several recent studies have found extended gaseous
halos of edge-on galaxies to be lagging in rotational speed when
compared to the midplane gas \citep{Heald06b,Heald07,Fraternali06}.
These offplane lags have been detected in both the diffuse ionized gas
(DIG) and HI.  There is some difficulty in comparing our measurements
of longslit rotation curves to other detailed measurements of offplane
gas which typically utilize 2-d information from radio
\citep{Barbieri05,Fraternali06}, Integral Field Units
\citep{Heald06b,Heald07}, and Fabry-Perot spectra \citep{Heald06} all
of which detect gas at larger scale-heights than those probed with our
offplane measurements.  The other major difference between these
previous studies and our offplane rotation curves is that we have
targeted lower mass galaxies.  The studies cited above target galaxies
with $220 > V_{max} > 110$ \kms\ while the sample studied here extends
to galaxies with rotation speeds of less than 60 \kms.

The gaseous lags observed in other systems are usually modeled with
either a galactic fountain that ejects gas to large scale-heights or
with a gas infall model where galaxies slowly accrete rotating gas.
The lack of significant lags in our H$\alpha$ rotation curves could
simply be a sign that these galaxies are not as active in forming
galactic fountains or accreting gas as the more massive galaxies.

\subsection{Ca {\sc ii} Rotation Curves}\label{Srcs}

To derive absorption line rotation curves, we require higher
signal-to-noise than for the \ha\ rotation curve.  We therefore sum
the 2D spectra in the spatial direction until the 1D spectra reaches
an adequate S/N ($\sim15$ per spectral pixel).  The resulting bins have
variable widths across the face of the galaxy, but roughly comparable
S/N per bin.  For the central regions of the galaxies the bin size is
around 10\arcsec\ while the outer regions and offplane components have
bin sizes $\sim20$\arcsec.  These bins correspond to $\sim$3-6 kpc at
the typical distances of the galaxies.  For reference, the typical
exponential disk radial scale lengths are $h_R\sim 12$\arcsec.

Extracting kinematic information from this data required developing a
new procedure.  In \citet{Yoachim05}, we tried both direct
$\chi^2$-fitting of a template spectrum as well as cross-correlation
of the galaxy with a stellar template to measure the stellar rotation
and line-of-sight velocity dispersion (LOSVD).  We have since
concluded that these traditional methods are not optimal for our data.
Direct fitting of a template star results in the template being
over-broadened (i.e., the fitted LOSVD diverges to large values).
This can be understood as the template star fitting the continuum
region of the galaxy spectrum at the expense of a small portion of the
absorption line.  Because the normalized continuum is very low S/N, it
is best fit by a straight line, which is equivalent to a stellar
spectrum which has been smoothed by a very broad filter.  In
\citet{Yoachim05} we were forced to hold the velocity dispersion fixed
during the $\chi^2$ minimization to prevent this problem.
Cross-correlation is also problematic, as the bright sky lines leave
regions of very low S/N and systematic residuals caused by variations
in the slit-width (Figure~\ref{sky_sub}).  Without a constant S/N
throughout the spectra, the cross-correlation peak can become skewed
by noisy regions.

To extract both velocity and velocity dispersion information from our
spectra we developed a modified cross-correlation technique that
allows regions of very low signal-to-noise to be masked.  This
modification prevents us from using the usual mathematical
techniques involving Fourier transforms and instead utilizes a
brute-force methodology.  What it lacks in mathematical elegance, our
procedure makes up for in functionality by being the only procedure we
know of that works on spectra that are both low S/N and contaminated
with systematic residuals.  We describe our modified
cross-correlation in detail in Appendix~\ref{ap1} and compare its
results to more traditional analysis methods in
Figure~\ref{cc_example}.  It may also be possible to use a penalized
pixel-fitting technique to measure the kinematics from our spectra,
but simulations show that the fitted parameters can become biased when
the S/N is low (60), or the LOSVD is poorly sampled
\citep{Cappellari04}.

For the stellar template, we used a KIII spectrum of star HD4388
downloaded from the Gemini archive along with accompanying calibration
frames of program GN-2002B-Q-61.  The stellar spectrum was reduced and
extracted using the Gemini IRAF routines.  Once extracted, the 1D
stellar spectrum was broadened with a Gaussian kernel to match the
instrumental resolution of our observations.  We found no significant
changes when trying different template stars and find our
uncertainties are never dominated by template mismatch.

%\clearpage
\begin{figure*}
\epsscale{.75}
\plotone{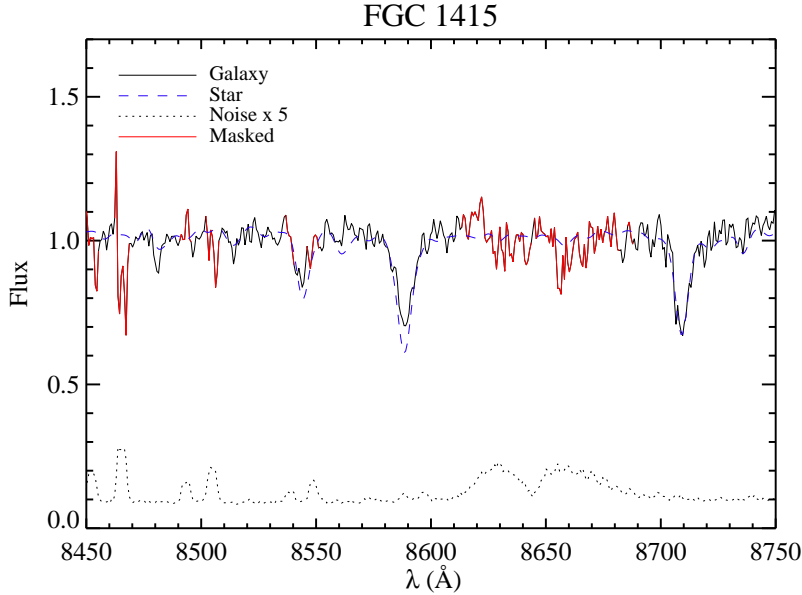}
\caption{An example of our extracted galaxy spectra.  The solid line
shows the normalized galaxy spectrum.  Red regions mark where the
spectra was masked due to sky line contamination.  The noise spectrum
(multiplied by 5) is plotted as a dotted line.  The blue dashed line
shows the best fit shifted and broadened stellar
spectrum.\label{ex_spec}}
\end{figure*}
%\clearpage

Because we have modified the traditional cross-correlation technique,
we have no formal means of calculating uncertainties in our fitted
velocity and LOSVD.  We therefore run a series of Monte Carlo
realizations to quantify the errors in our fitting procedure.  For
each galaxy, we create 100 artificial 2D spectra.  A template stellar
spectrum is shifted to match a realistic rotation curve, and broadened
to simulate both stellar velocity dispersion and instrumental
resolution.  We vary the detailed shape of the rotation curve and
velocity dispersion for each realization by $\sim20$\%.  The fake
spectra have radial exponential flux profiles similar to the real
galaxies.  We add Poisson noise to the artificial spectra, as well as
systematic residuals by adding regions of sky from our science frames
that do not have any detectable objects.  Thus, our artificial spectra
have both the same Gaussian sky background and similar systematic
residuals as the real data.

Once the artificial spectra are made, we extract and analyze 1D
spectra identically to the real data (i.e., we use the same extraction
windows and the cross-correlation with masking procedure).  In many
instances, we found that our measured LOSVD poorly matched the input.
The loss of reasonable LOSVD measurements is dominated by how many of
the Ca{\sc ii} lines are masked due to sky line contamination.  We
therefore clip points where the Monte Carlo error analysis suggests we
cannot reliably recover the input parameters (i.e. the RMS error
between input and output is $>50$ \kms\ or the output has a systematic
error of $>20$ \kms).  These clipped regions typically correspond to
regions of the rotation curve where the Ca triplet line passes through
a large sky residual.

Our final extracted rotation curves, LOSVDs, and Monte Carlo derived
uncertainties are plotted in Figure~\ref{all_rc} along with $R$-band
images of the galaxies showing the Gemini longslit placements.  

%\clearpage
\begin{figure*}
\epsscale{.5}
%\plottwo{f4a.eps}{f4b.eps} 
%\plottwo{f4c.eps}{f4d.eps}
%\plottwo{f4e.eps}{f4f.eps}
%\plottwo{f4g.eps}{f4h.eps}
%\plotone{f4i.eps}

\plotone{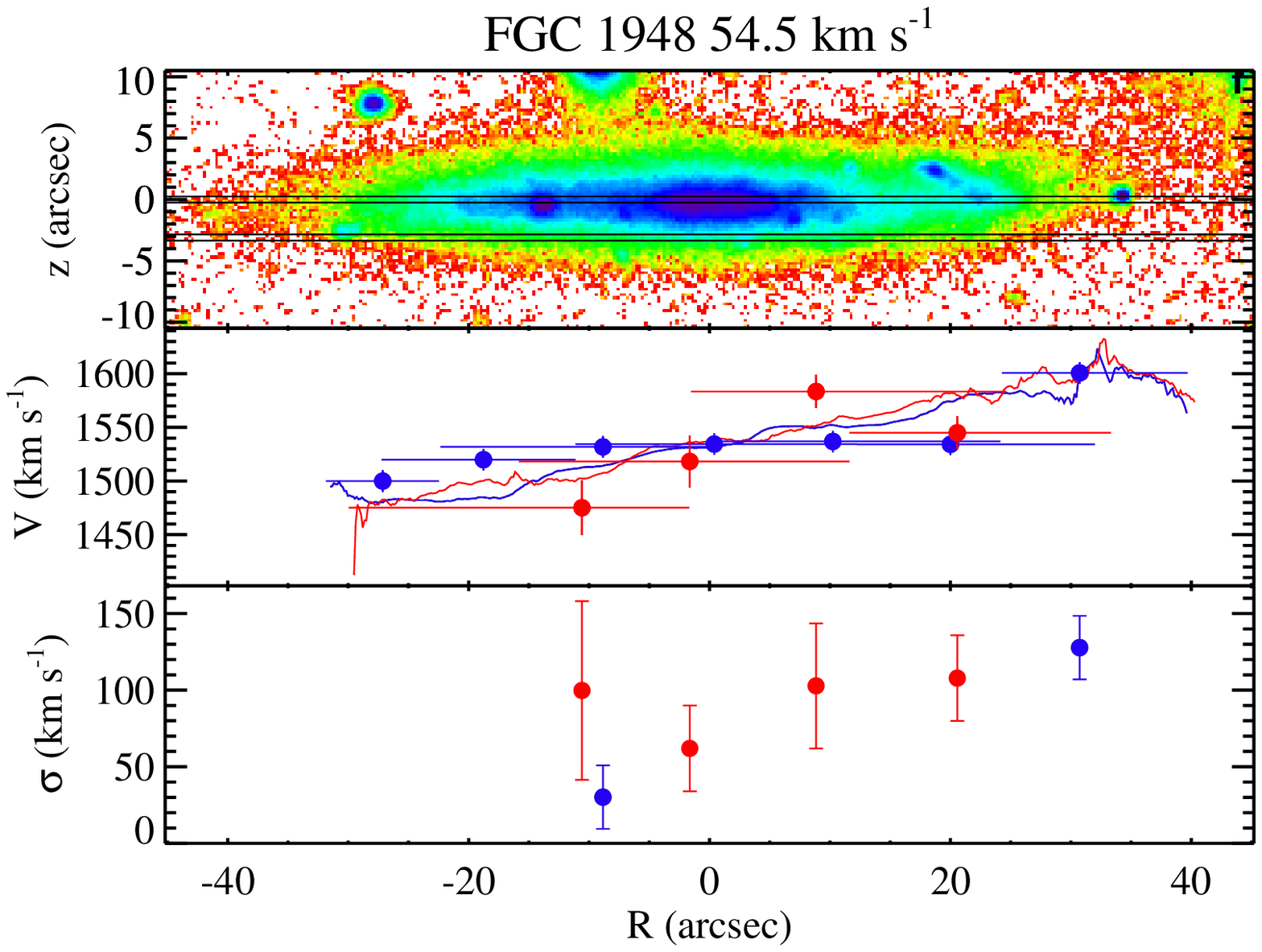}
\plotone{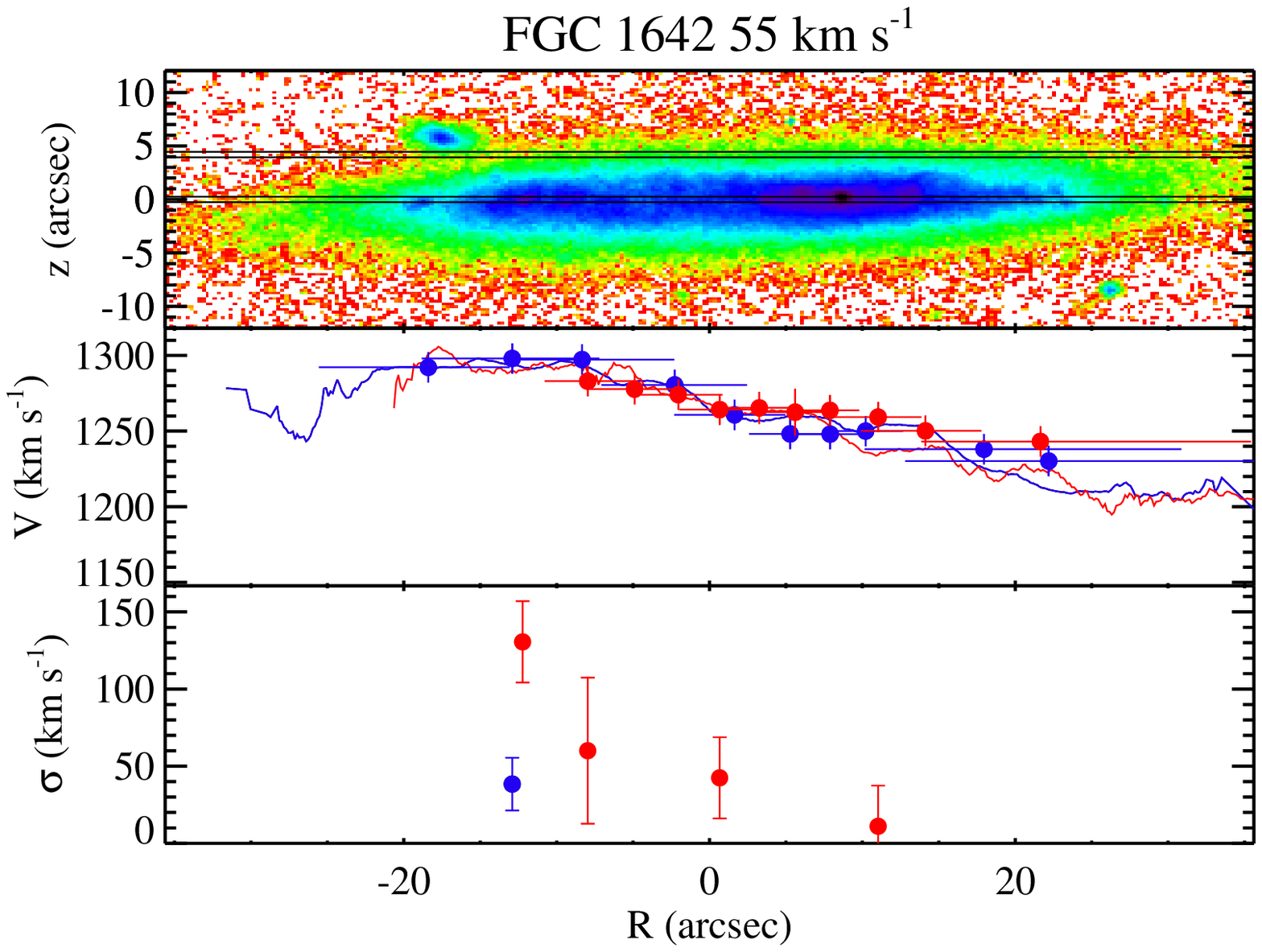}
\plotone{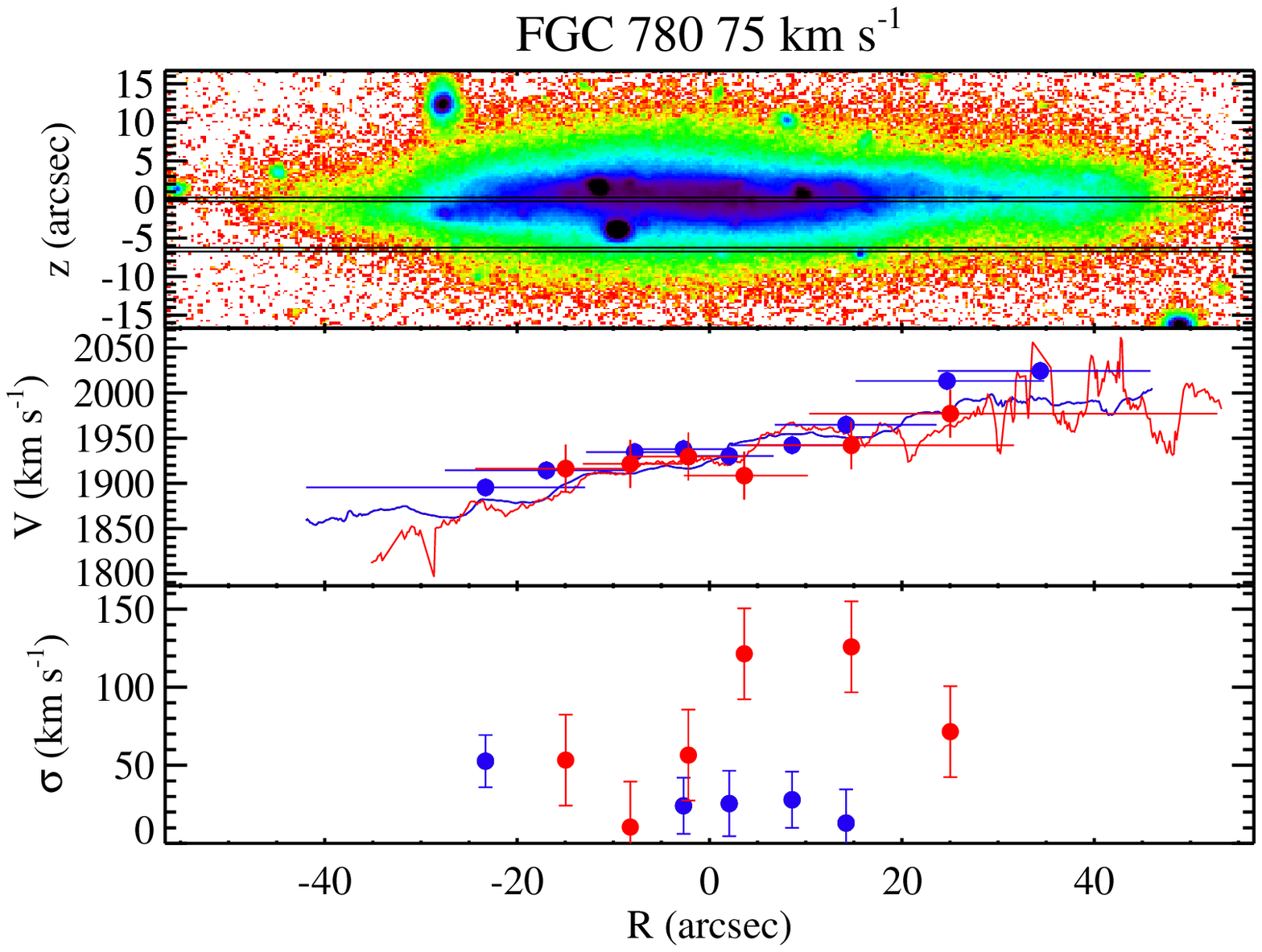}
\plotone{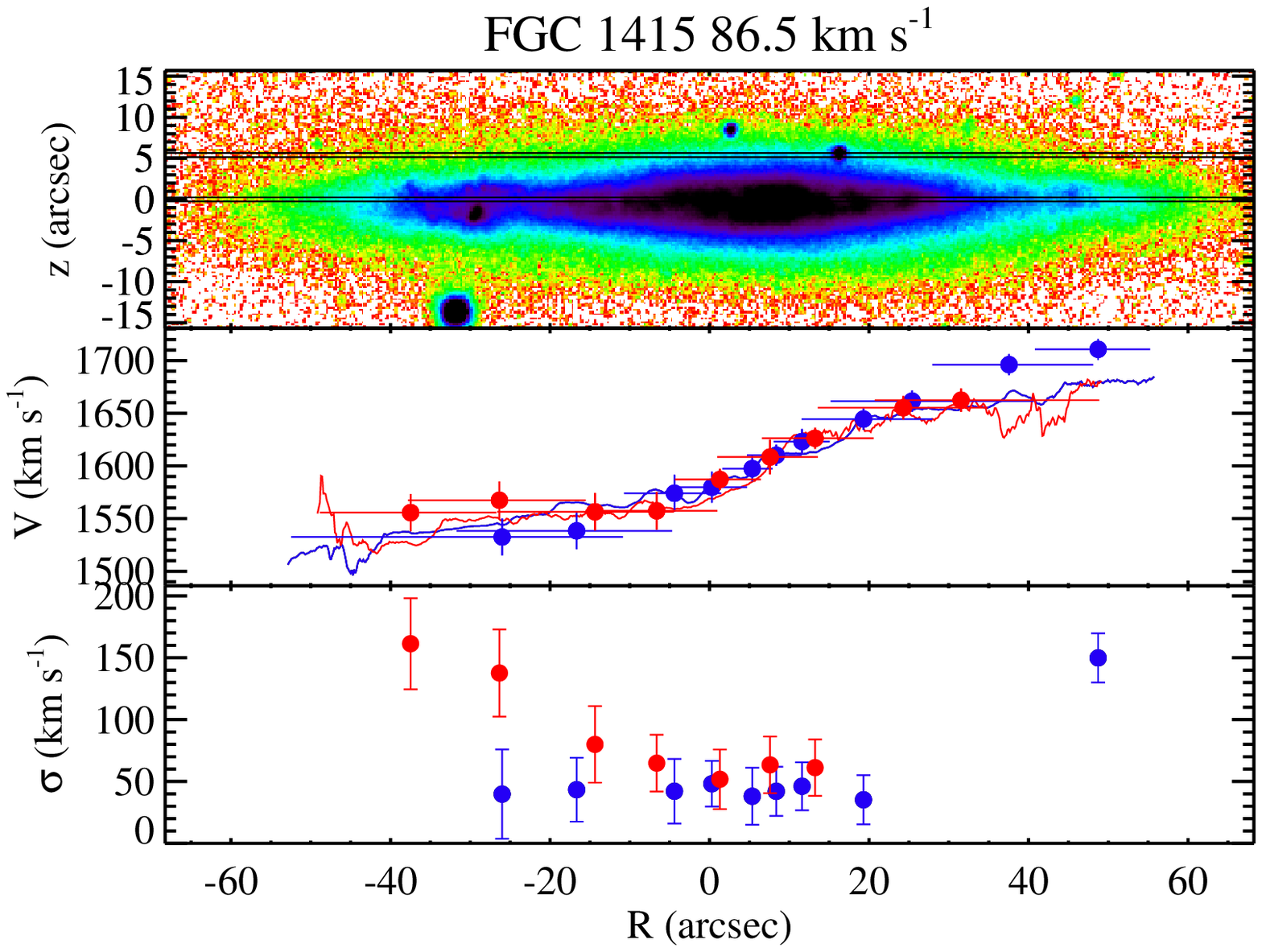}
\plotone{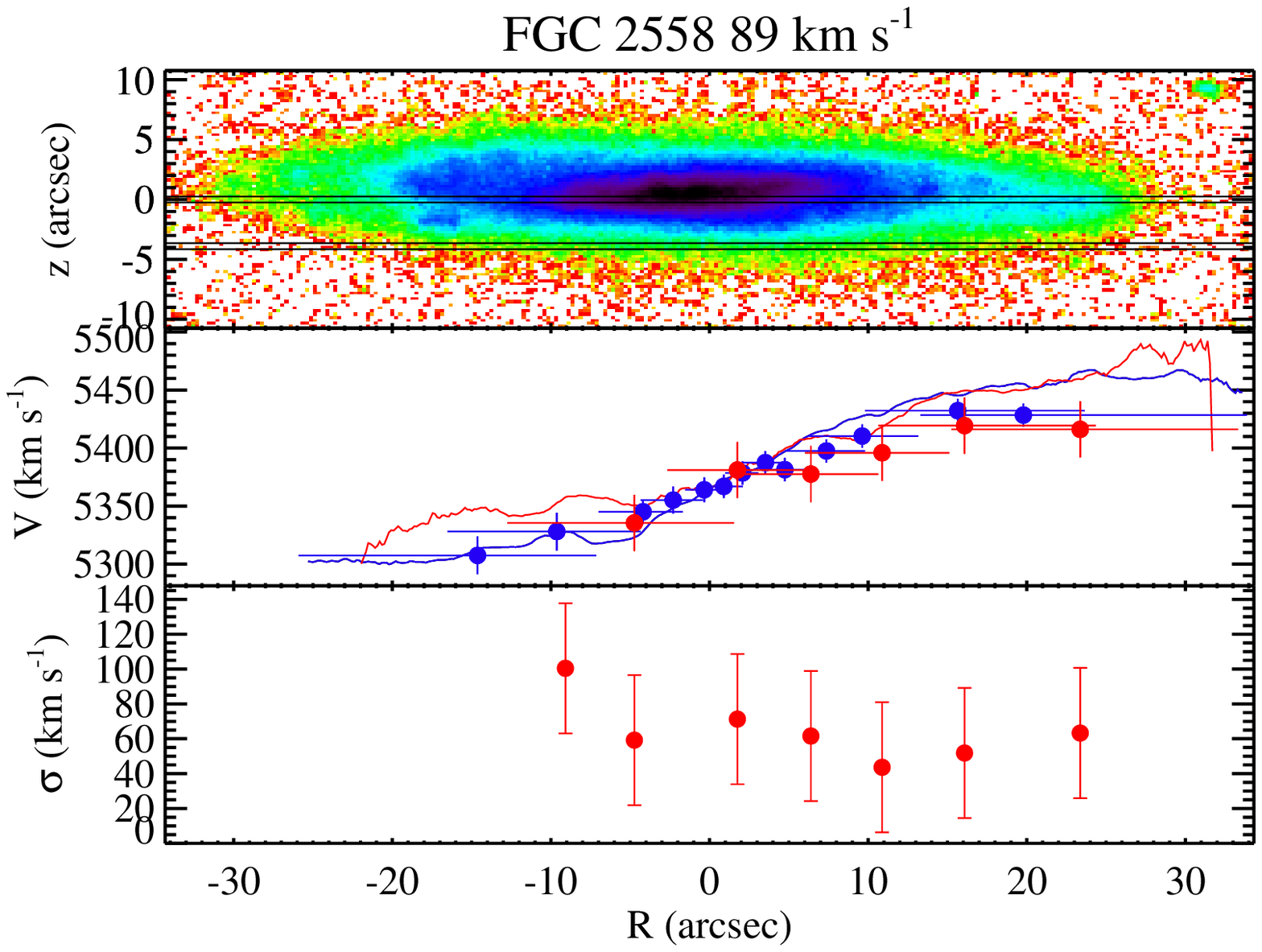}
\plotone{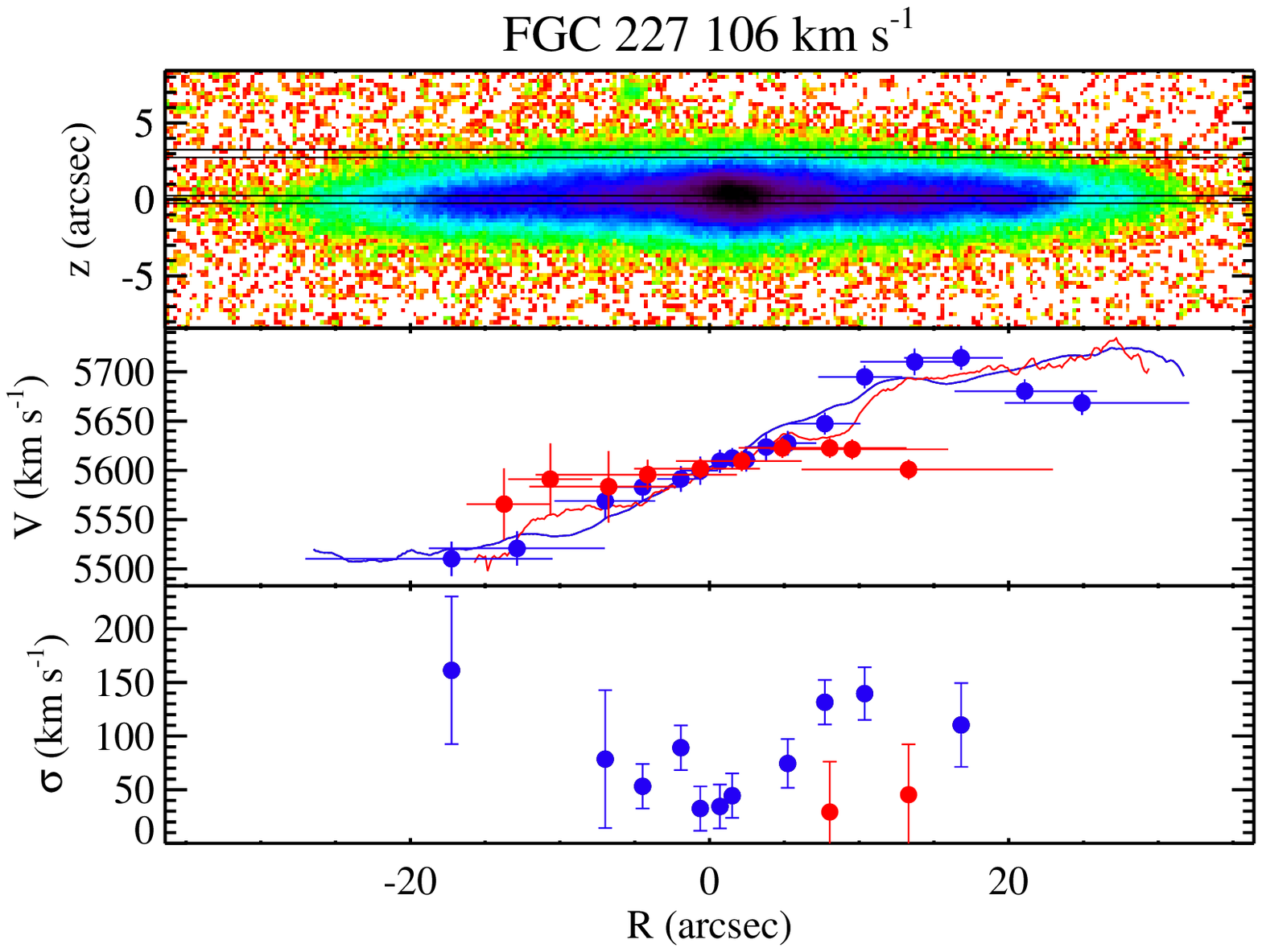}

\caption{Rotation curve measurements for each galaxy.  Top panels:
$R$-band images of each galaxy.  The color scale goes from dark blue
($\mu_R=21$) to green($\mu_R=23$), to red/white ($\mu_R=25.5$).  Solid
black lines have been drawn where the Gemini long-slit jaws were
placed.  Middle Panels: Rotation curves for midplane (blue) and
offplane (red).  Points with error bars are from Ca {\sc ii}
measurements.  Vertical error bars are uncertainties derived from
Monte Carlo simulations, horizontal error bars show the spectral
extraction regions.  Small lines show velocities measured from the
H$\alpha$ emission lines.  Bottom Panels: Stellar velocity dispersions
measured from the Ca {\sc ii} feature.  All error bars are from a
Monte Carlo simulation.  Points with overwhelmingly large error-bars
or large systematic uncertainties have been omitted.  \label{all_rc}}
\end{figure*}

%\begin{figure*}
%\figurenum{4}
%\epsscale{.5}
%\plotone{f4d.eps}
%\plotone{f4e.eps}
%\plotone{f4f.eps}
%\caption{ \emph{continued}.}
%\end{figure*}

\begin{figure*}
\figurenum{4}
\epsscale{.5}
\plotone{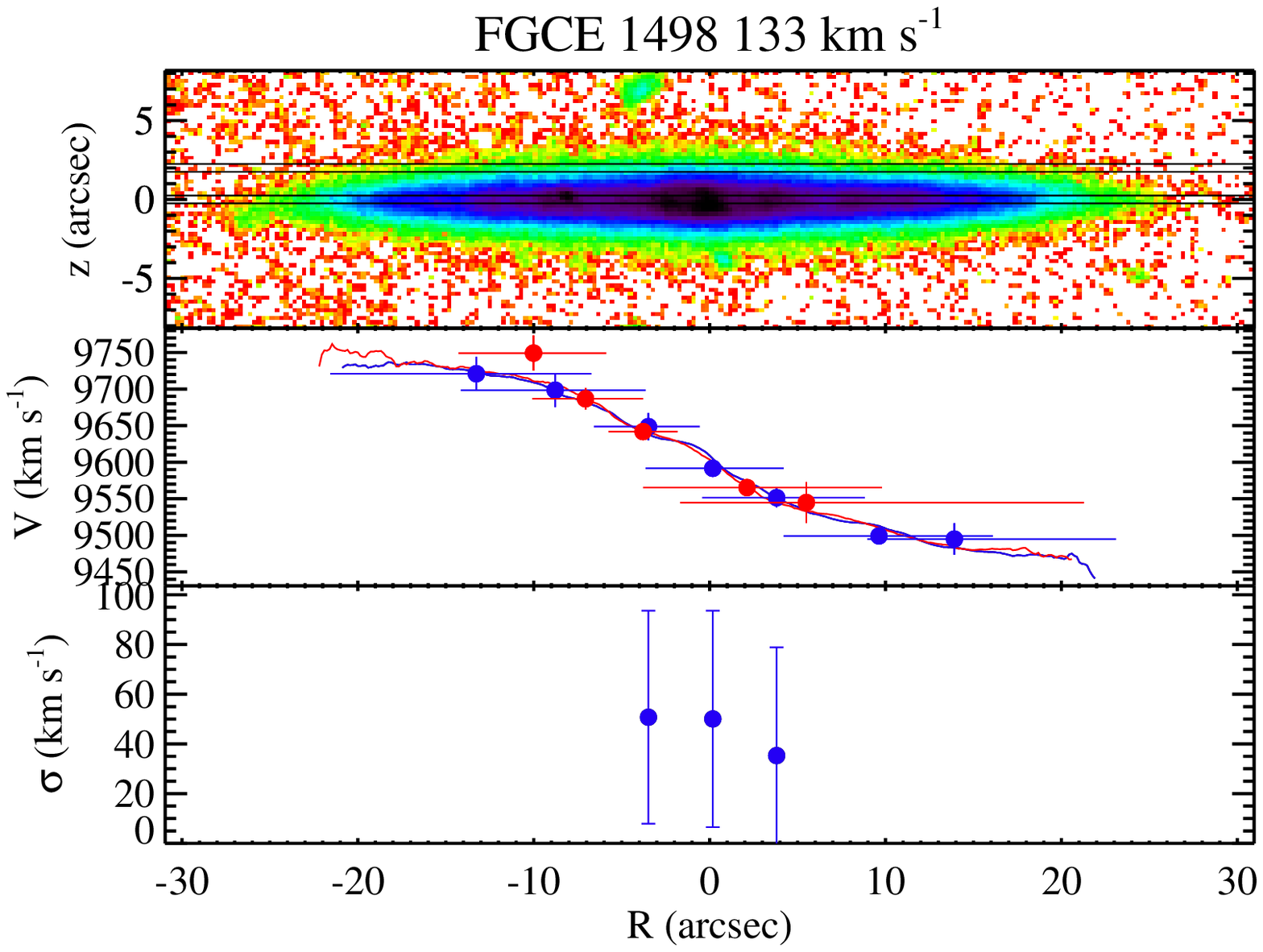}
\plotone{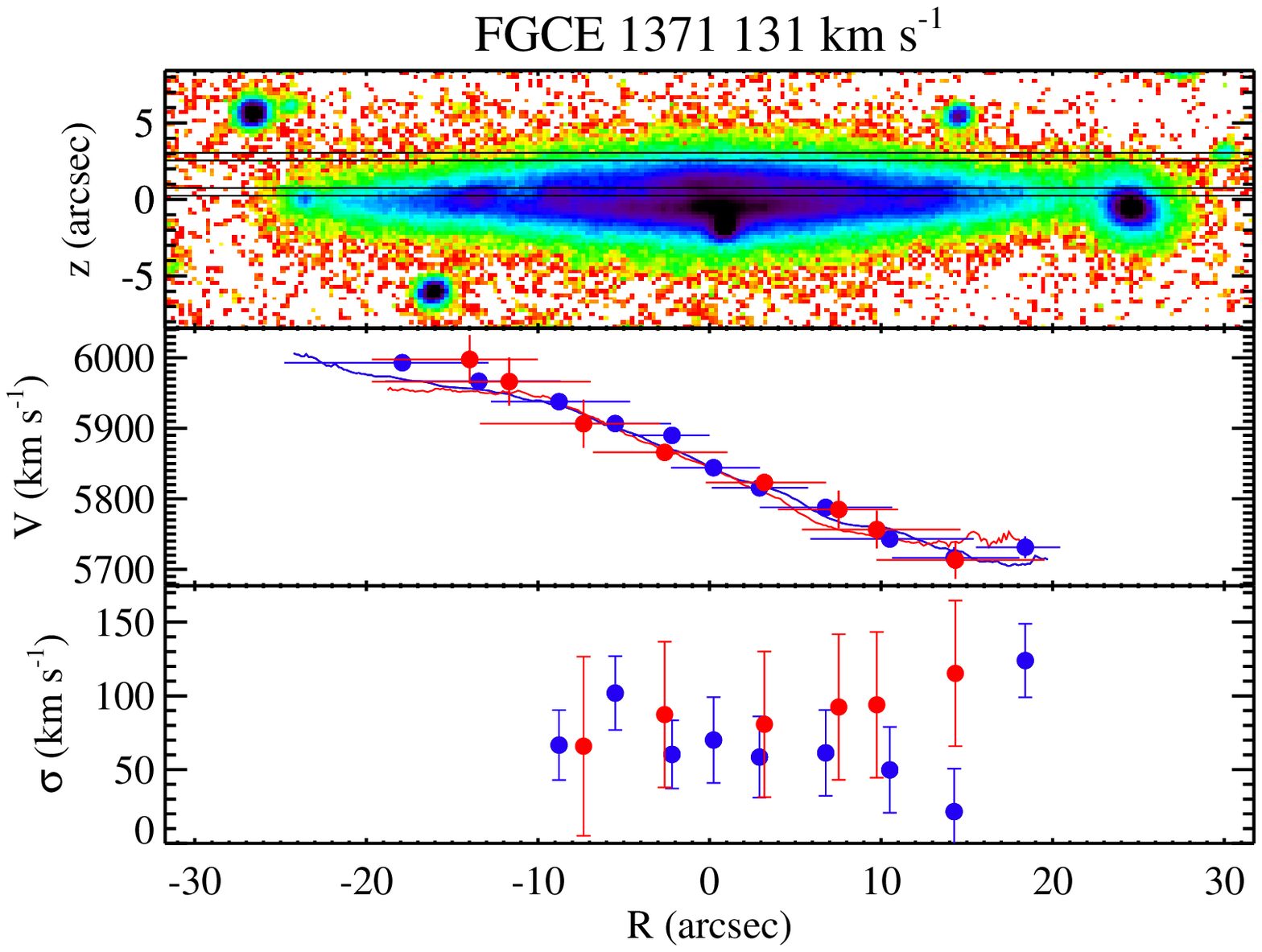}
\plotone{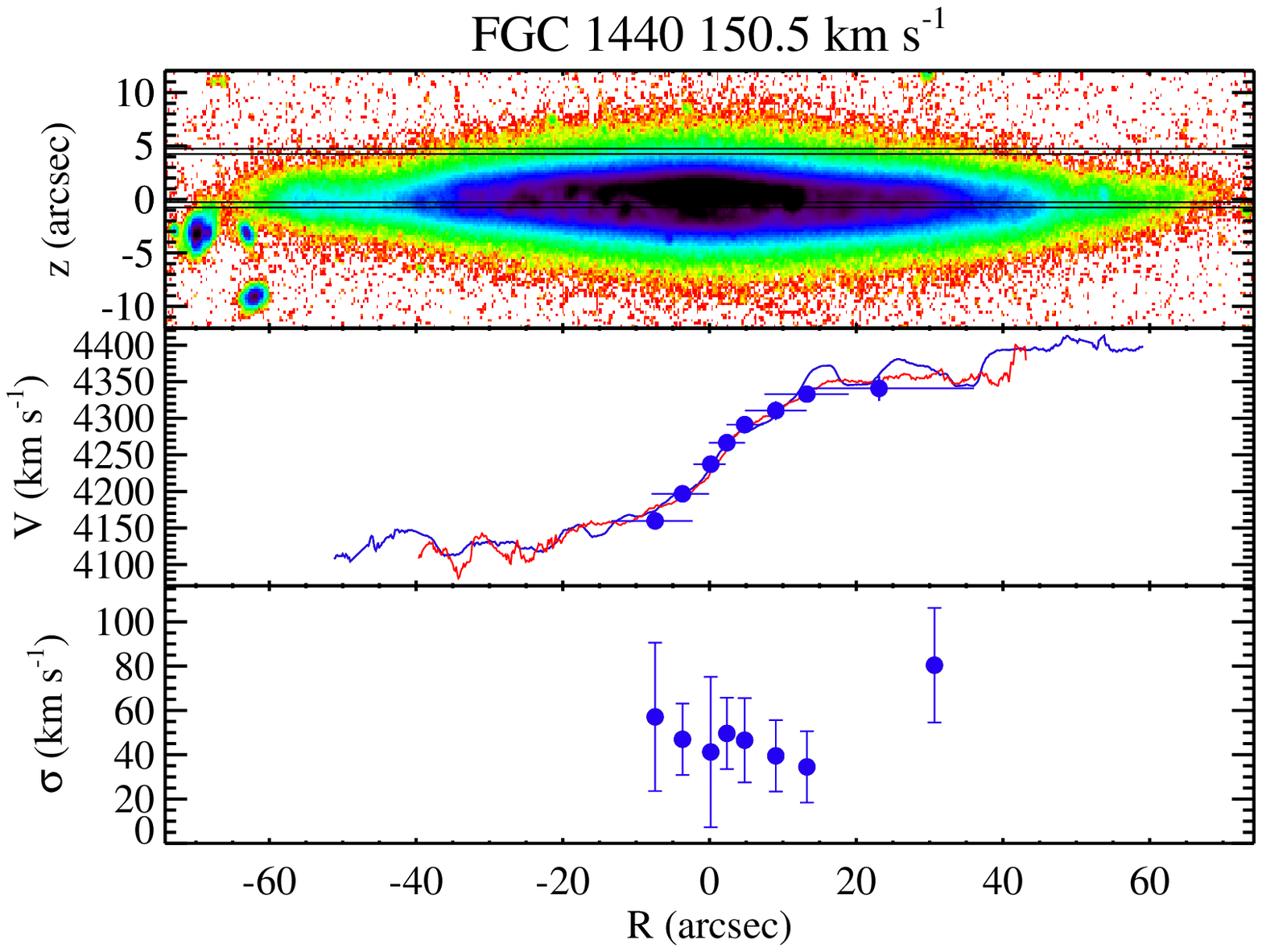}
\caption{\emph{continued}.}
\end{figure*}
%\clearpage

\section{Stellar Kinematics}\label{Sstell_kin}

Although we attempted to place our slits in regions of the galaxies
where the thin and thick disk light makes up the majority of the flux,
it is nearly impossible to target regions where one stellar
component completely dominates the flux.  In the lower-mass galaxies,
we found that the thick disk is a major stellar component and we
should expect spectra taken along the midplane to include a large
amount of thick disk light.  In the higher mass galaxies, the thin
disk is the dominant component, and we are forced to observe off-plane
regions that still contain a large fraction of thin disk light.  Using
the photometric fits of \citet{Yoachim06}, we can estimate the
fractional flux levels of the thin and thick disk at each slit
position.  Because each slit position should include both thin and
thick disk stars, we make an attempt to model the true underlying
rotation curves for each population.  For simplicity, we assume that
the thin and thick disk stars are each rotating cylindrically and
therefore have the same rotation curve for both the on and off-plane
observations.  We discuss this choice in more detail in \S\ref{ex_lags}.

The details of the vertical profiles of the stellar disks (exponential
vs sech$^2$) can dramatically influence what fraction of the midplane
light belongs to thin disk stars versus thin disk stars.  As in
\citet{Yoachim05}, we adopt a series of photometric decomposition
models that should cover the full range of possible thin and thick
disk fractions.  At one extreme, we use a simple model where we assume
the midplane is composed of only thin disk light and the offplane
observations purely thick disk stars.  As a more accurate model, we
use the thin/thick fractions from the best fitting models of
\citet{Yoachim05} as well as models where we vary the parameters by
their 1-$\sigma$ values to create a ``bright-thick and faint-thin''
model along with a ``faint-thick and bright-thin'' model.  The
differences between the thin and thick disk scale lengths are small
enough that we do not expect much radial variation in the fraction of
thin and thick disk light.

In \citet{Yoachim05}, we fit analytic functions to the stellar
rotation curves to decompose the thin and thick disk components.  This
worked well for the initial two galaxies we observed, but our expanded
sample now includes galaxies with slightly irregular kinematics that
are not well described by common parameterizations of rotation curves.
Instead of using an analytic function, we use the midplane H$\alpha$
rotation curve as a basis function for the overall shape of the
rotation curve.  Because we are most interested in finding the
velocity of the thin and thick disk stars relative to each other, we
compare them both to the well resolved and high signal-to-noise
midplane H$\alpha$ rotation curve.  Using the \ha\ rotation curve
reduces the number of parameters that need to be fit to characterize
the stellar rotation curves.

We model the stellar rotation curves as $V_{stars}(R)=x
V_{\rm{H}\alpha}(R)+c$.  We constrain $c$ to be in the range $\pm5$ (to
account for any small error in wavelength calibration between frames
or regions on the chip) and $x$ is limited to $-1 < x < 1.4$, allowing
for stars to be rotating faster than the gas by up to 40\% ($x$=1.4),
not rotating ($x$=0), or counter-rotating with the opposite velocity of
the H$\alpha$ ($x$=-1).

\begin{figure*}
\epsscale{.4}
%\plottwo{f5a.eps}{f5b.eps}\\
%\plottwo{f5c.eps}{f5d.eps}\\
%\plottwo{f5e.eps}{f5f.eps}\\
%\plottwo{f5g.eps}{f5h.eps}
\plotone{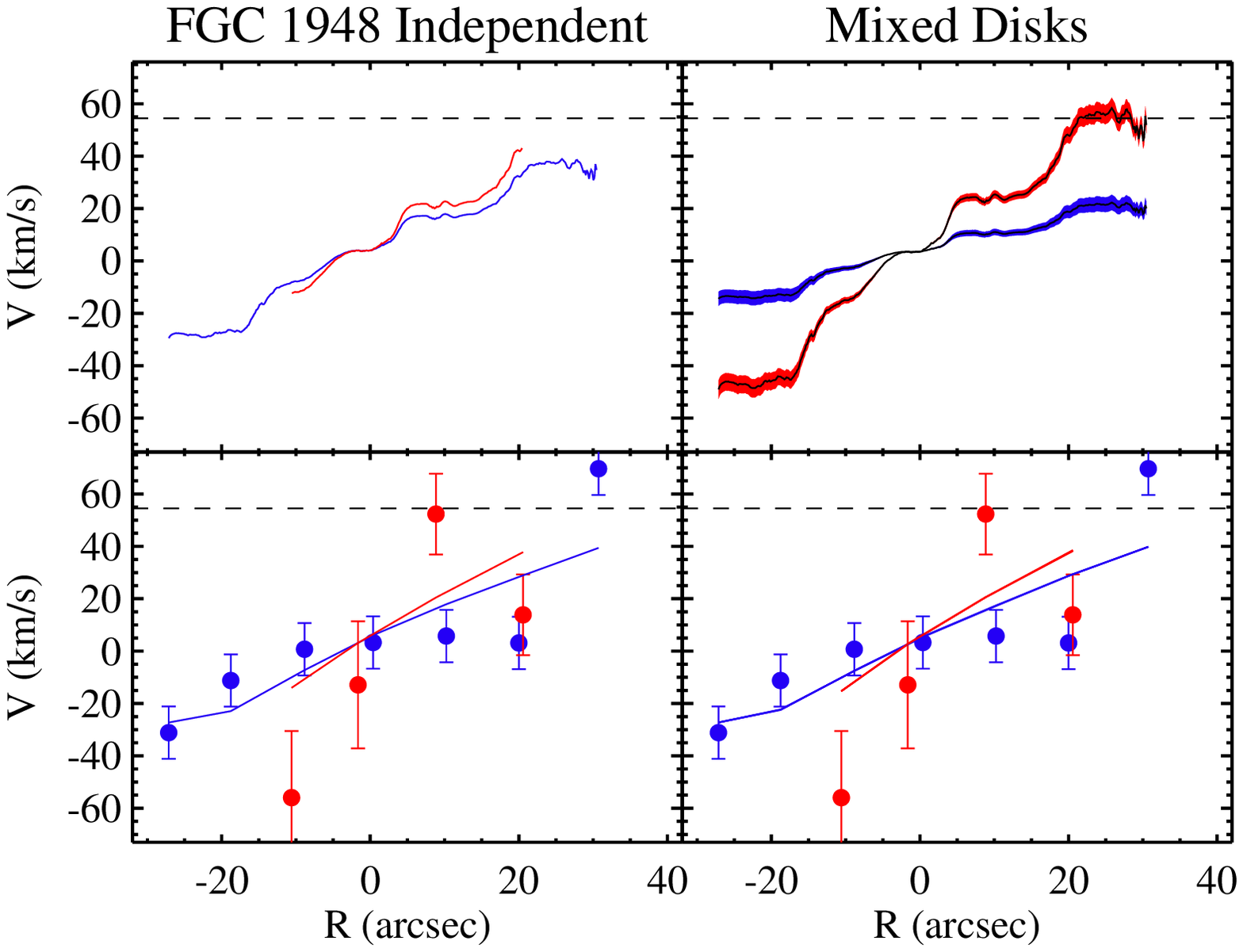}
\plotone{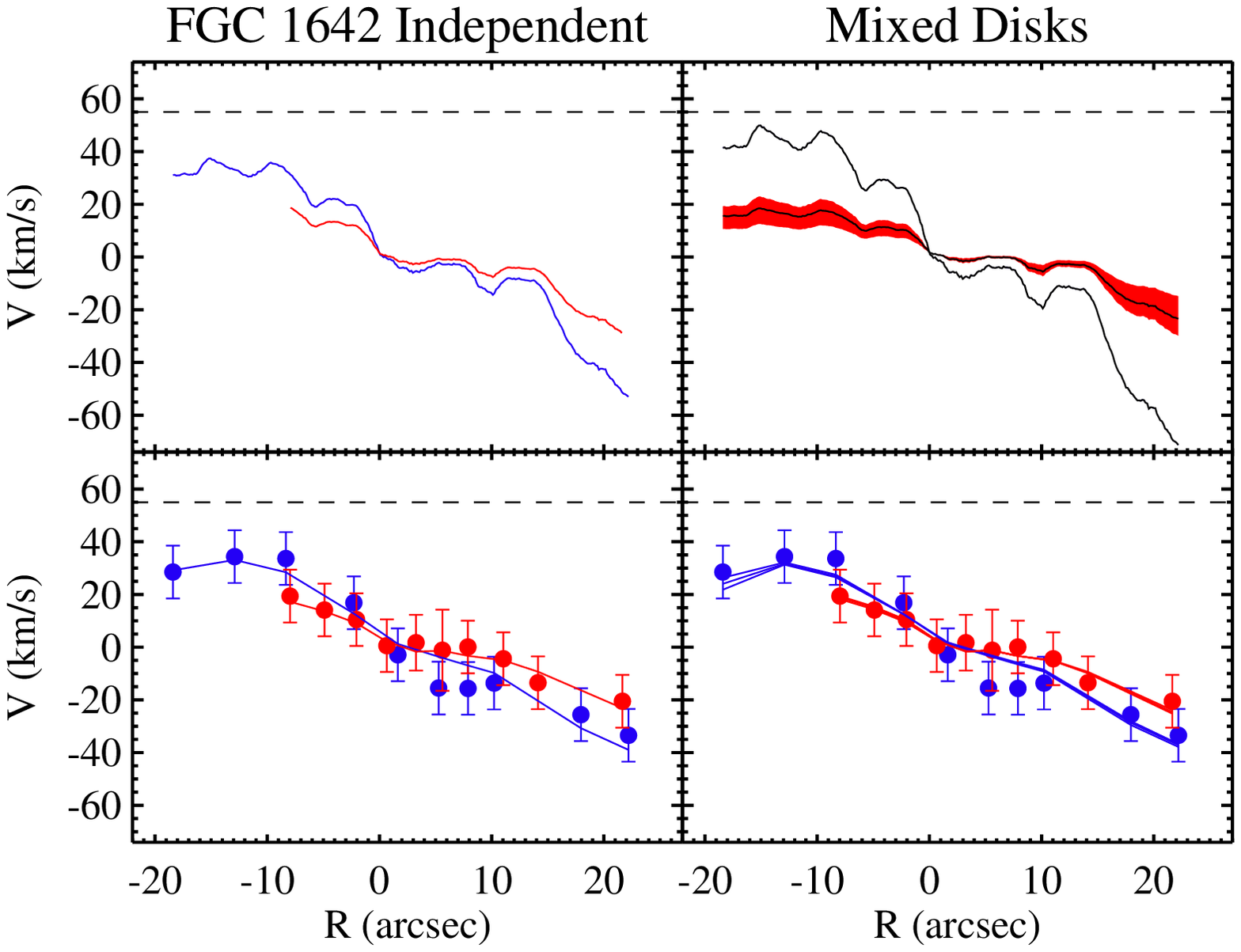}
\plotone{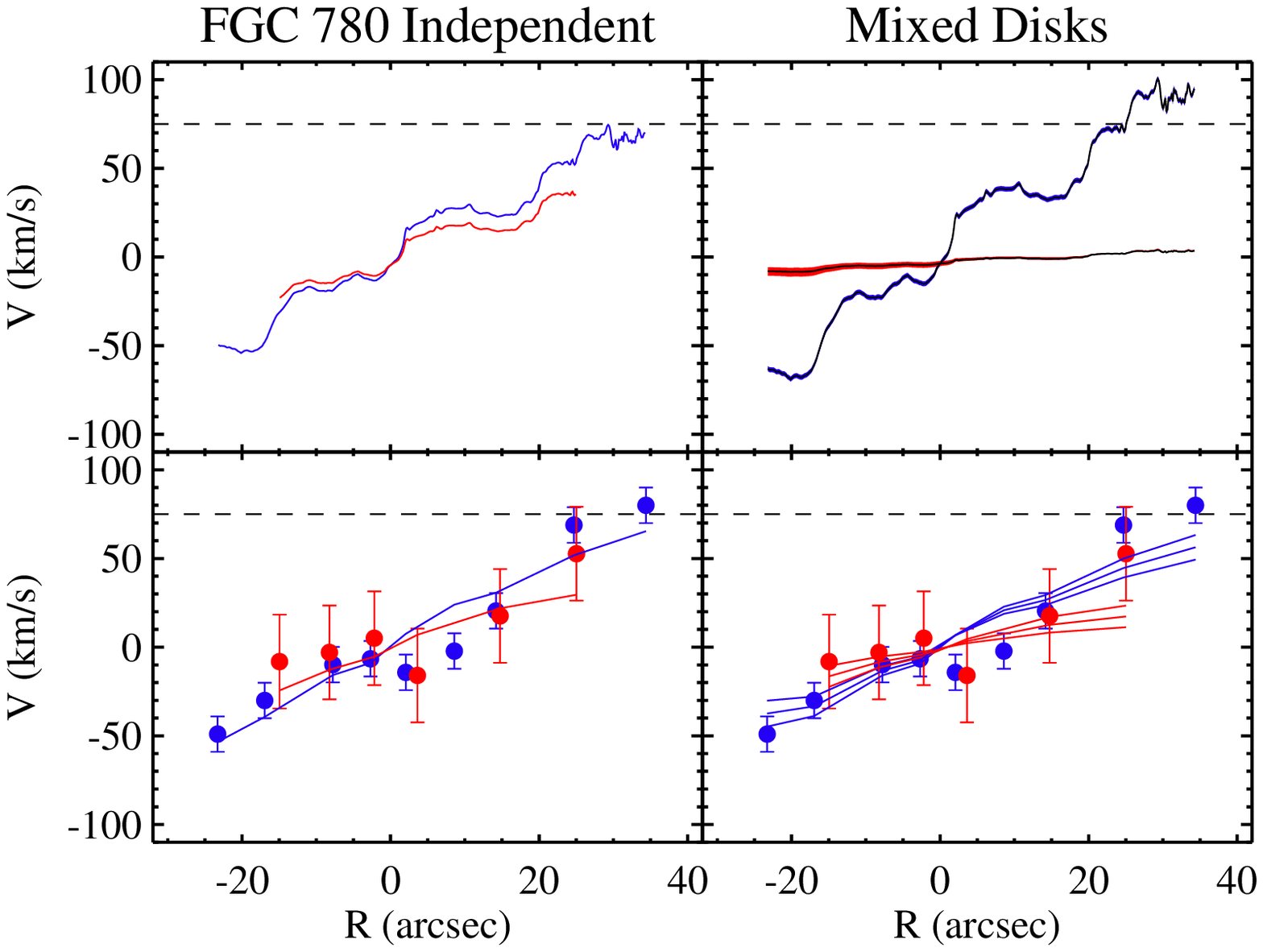}
\plotone{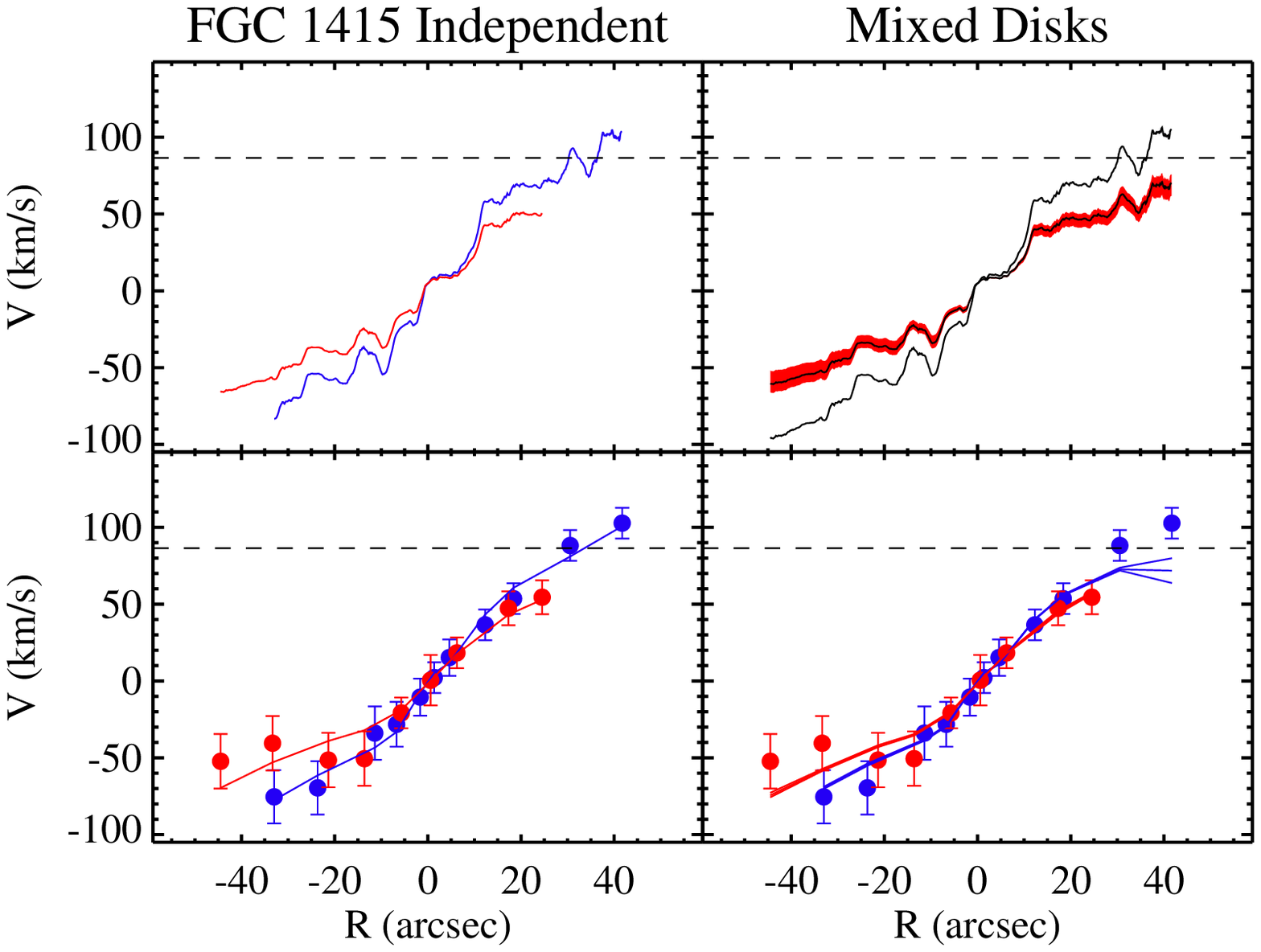}
\plotone{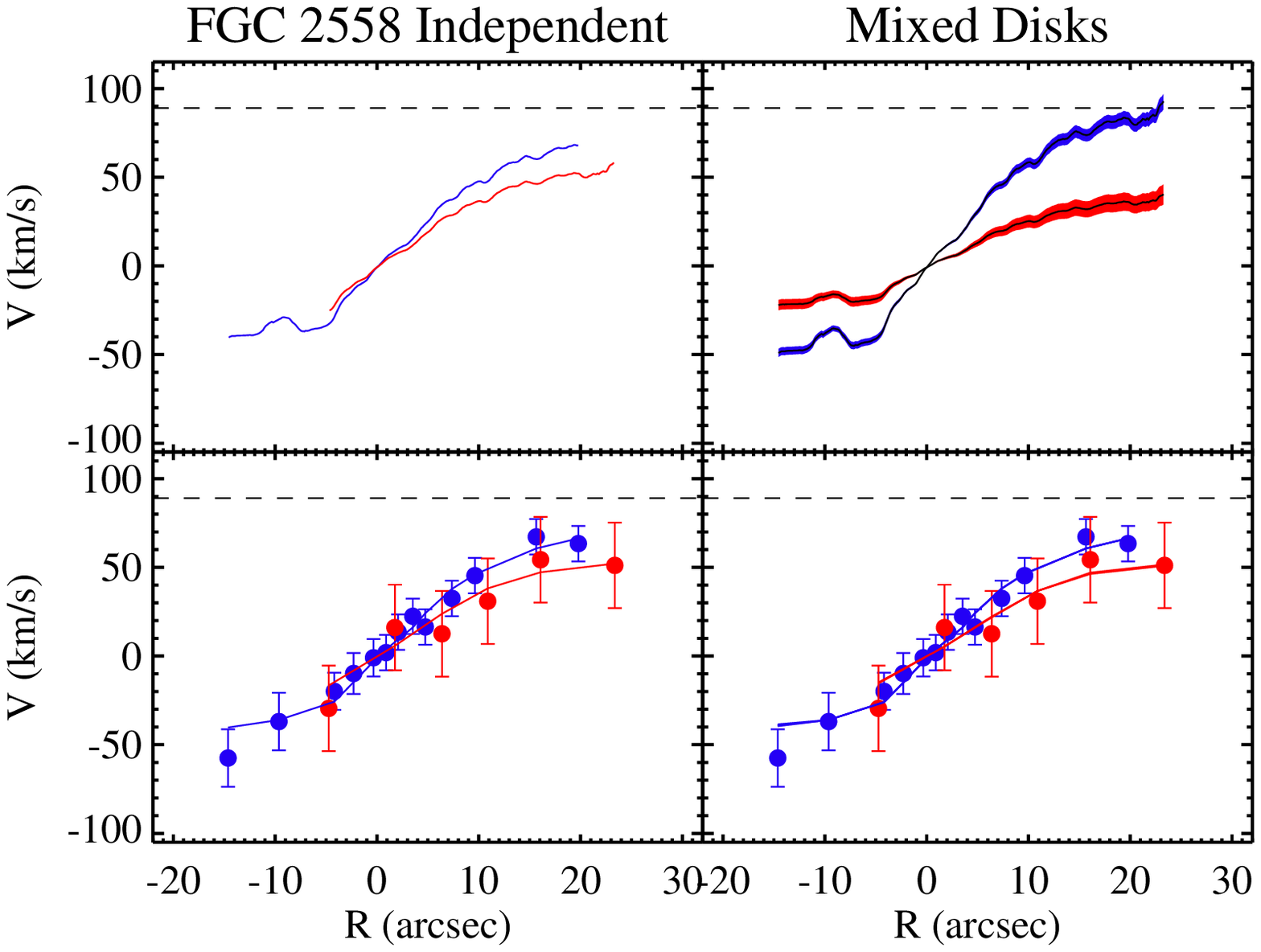}
\plotone{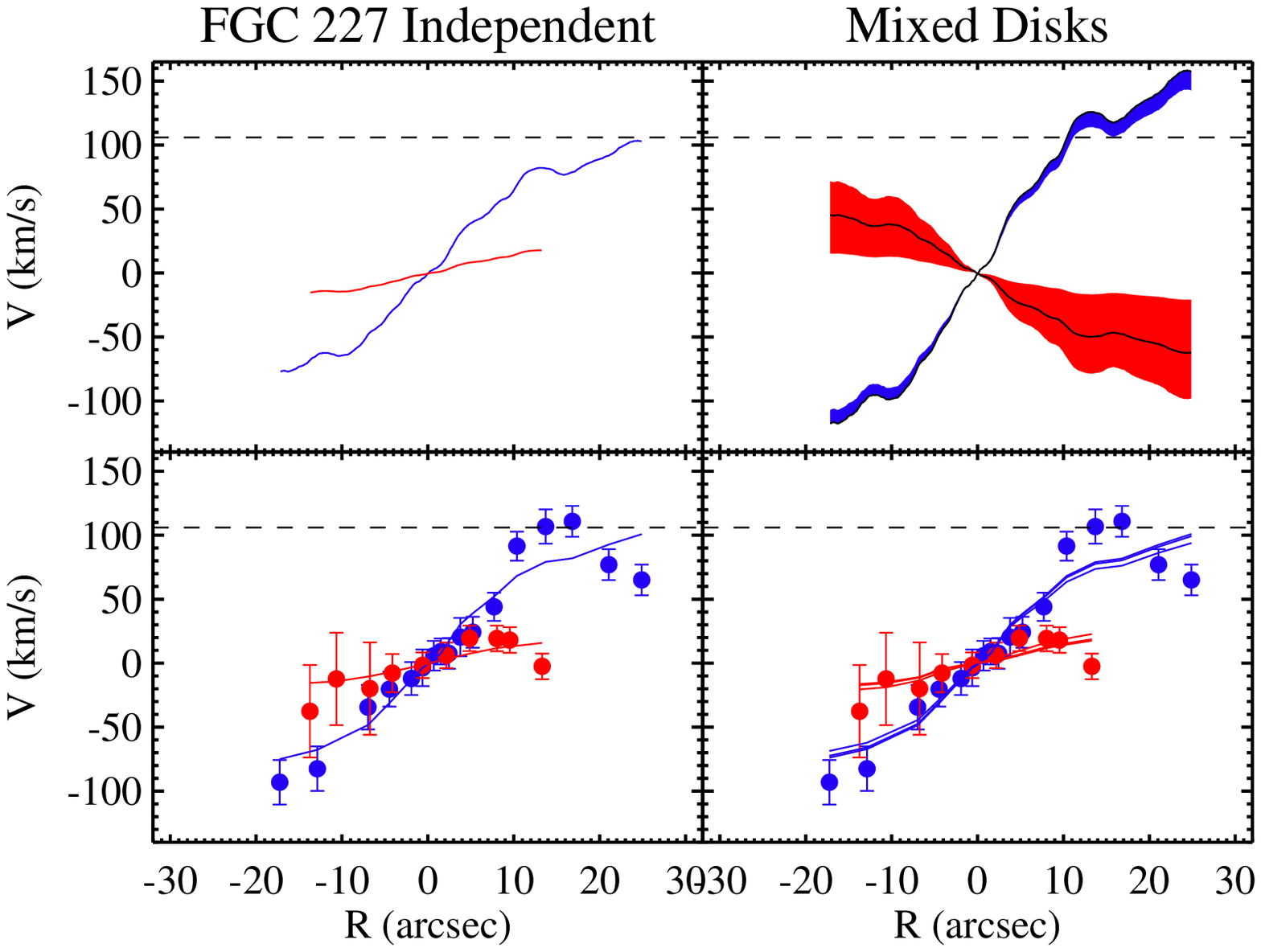}
\plotone{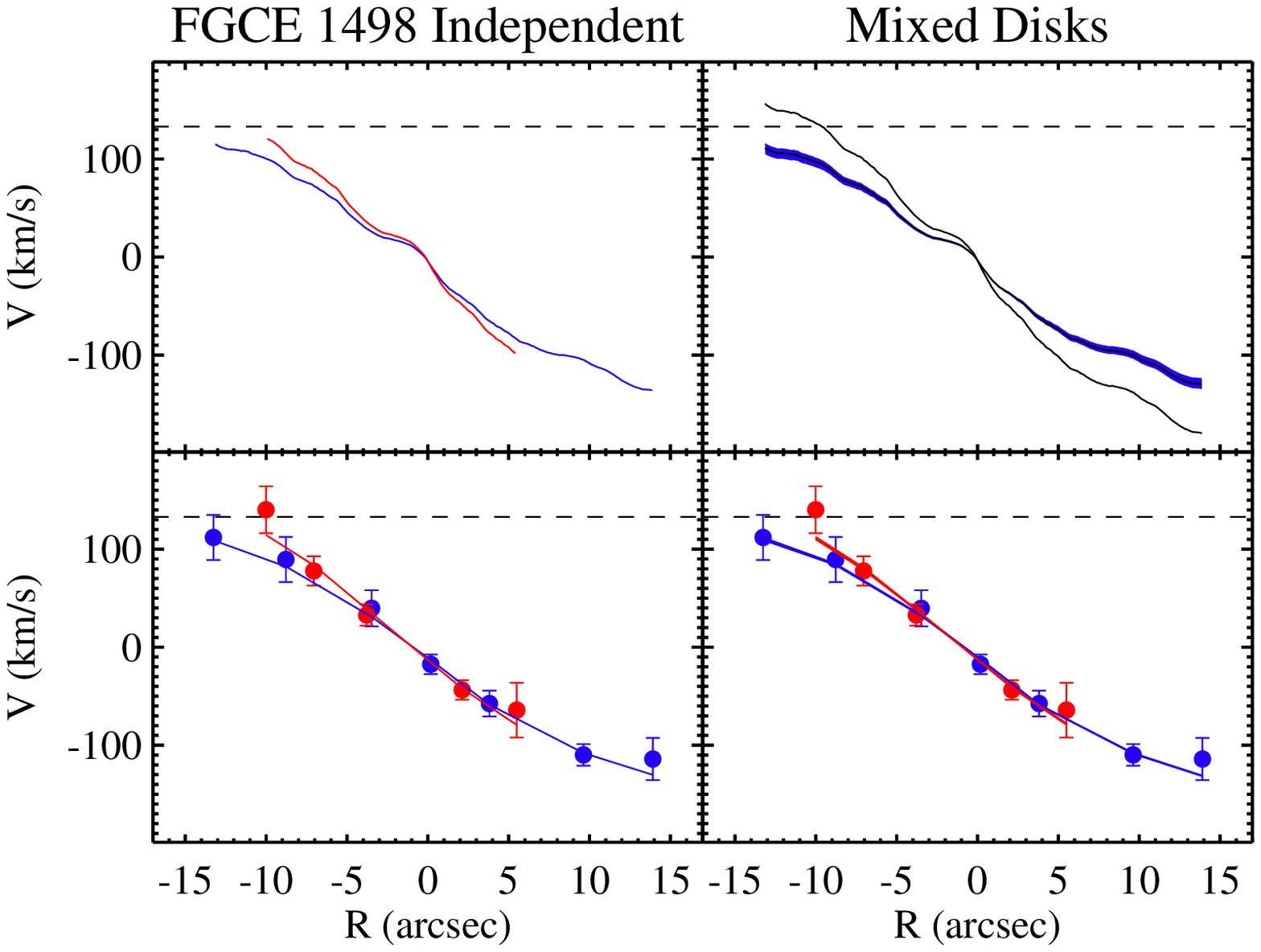}
\plotone{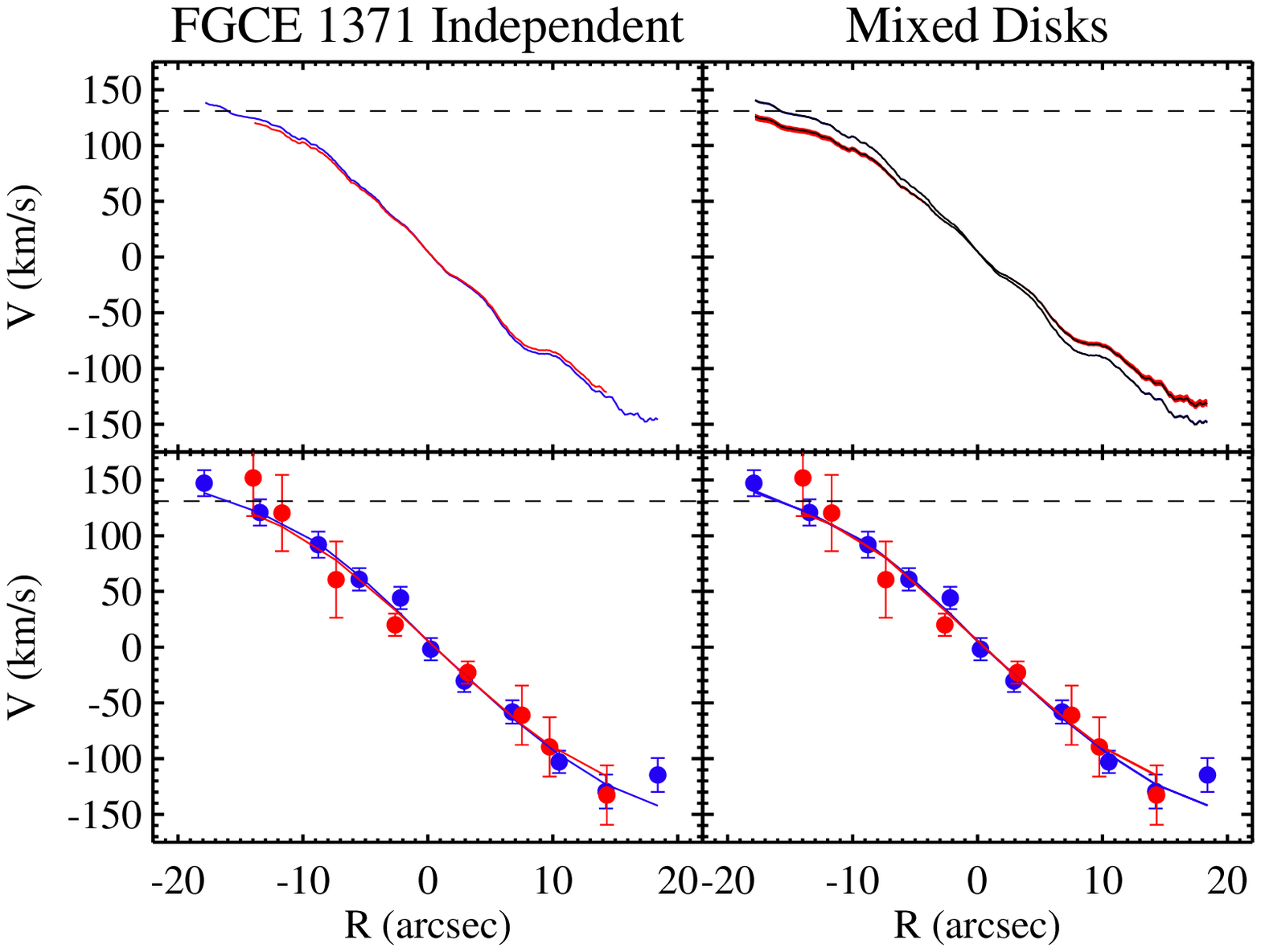}

\caption{The results of fitting various rotation curve models to our
data.  The top left panels show fits of the simple model where the
midplane and offplane observations are fit independently.  Upper right
panels show shaded regions show the range of fits derived from varying
the fraction of thin and thick disk light at each slit position.
Solid lines show the fits for when we use the thin and thick disk
fractions of the photometric fits in \citet{Yoachim06}.  Lower panels
show the observations as points and solid lines show the models from
the above panels once they have been flux weighted and binned in the
same manner as the observations.  Throughout, red is used for thick
disk/offplane and blue is used for thin disk/midplane.  Each panel has
a dashed line showing the W50/2 value from the literature.  FGC 1440
is not shown because we failed to measure a stellar rotation curve in
the offplane position.  \label{decomp_rc} }
\end{figure*}

The decomposed rotation curves are plotted in Figure~\ref{decomp_rc}.
The left hand panels show the best fit stellar rotation curve scaled
from the \ha\ at each slit position.  If there were no
cross-contamination of thin and thick disk stars, then the offplane
and midplane rotation curves would show the true thick and thin disk
kinematics.  The right hand panels show the more realistic case where
we have adopted likely amounts of thin and thick disk contamination at
each slit position before inferring the underlying kinematics of each
population.

For the higher mass galaxies, we find no substantial difference
between the thin and thick disk rotation curves, even when we correct
for the expected cross contamination.  There is a slight tendency for
the thick component to be lagging, but never by more than 5 \kms.
In the higher mass galaxies, we have therefore either failed to
observe an offplane region with a high enough thick disk flux
fraction, or the thick disks are not lagging significantly compared to
the thin disk in these systems.

For the low-mass galaxies, we find a wide range of behavior.  The fits
for FGC 1948 diverge, as the stellar rotation curves do not show
coherent rotation at either slit position.  For the rest of the
galaxies, the best fits find thick disks that are slightly lagging
compared to the thin (FGC 2558, FGC 1415), that are lagging to the
extent of near non-rotation (FGC 1642, FGC 780), and that are fully
counter-rotating (FGC 227).  We note that there is strong qualitative
agreement with initial results in \citet{Yoachim05} for FGC 1415 and
FGC 227.

%\clearpage

%\begin{figure*}
%\epsscale{.5}
%\figurenum{5}
%\plotone{f5d.eps}
%\plotone{f5e.eps}
%\plotone{f5f.eps}
%\caption{\emph{continued}.}
%\end{figure*}

%\begin{figure*}
%\epsscale{.5}
%\figurenum{5}
%\plotone{f5g.eps}
%\plotone{f5h.eps}
%\caption{\emph{continued}.}
%\end{figure*}
%\clearpage

\subsection{Velocity Dispersions}

The low signal-to-noise of our spectra prevents us from reliably
measuring velocity dispersions for many of our galaxies.  Most of the
galaxies with high quality spectra have very low velocity dispersions,
as we would expect from systems predominantly supported by rotation.
Given that our instrumental resolution is 60 \kms, we are unlikely to
resolve the line widths in galaxies where $\sigma/V_c<0.6$, for the
$V_c<100$ \kms\ galaxies that dominate our sample.  The major
exceptions are FGC 1948, which has an irregular rotation curve, and
FGC 227, which has a counter-rotating thick disk.

FGC 1948 has surprisingly large LOSVD, with many regions of the disk
having $\sigma>100$ \kms.  For comparison, most of the other galaxies
in our sample have LOSVDs across the disk of $\sim 50$ \kms,
essentially the same as our instrumental resolution at the Ca {\sc ii}
triplet.  The stellar rotation curve for FGC 1948 also shows large
deviation from the H$\alpha$ RC, suggesting that the stars in this
galaxy might not be fully rotationally supported and/or fully
dynamically relaxed.

FGC 227's LOSVD also deviates from the simple interpretation of a
dynamically cold rotating disk.  In the midplane observations, the
central regions of FGC 227 appear cold ($\sigma\sim40$ \kms), but the
outer disk reaches LOSVD values of 100-150 \kms.  This makes little
sense for a galaxy with a well defined rotation curve as the intrinsic
stellar velocity dispersions should be decreasing with radius.  In
contrast, the LOSVD can be well explained by a rotationally supported
galaxy if there are two stellar populations moving in opposite
rotational directions.  As our rotation curve decomposition showed,
FGC 227 is best fit by a model where the thick disk is
counter-rotating relative to the thin disk.  As we showed in
\citet{Yoachim05}, this would cause an increase in the observed
velocity dispersion of order 50 \kms.  Similar projection effects are
found in elliptical galaxies with counter-rotating cores as they also
show radially increasing LOSVDs \citep{Geha05}.

\section{How Much Counter Rotating Material Could There Be?} \label{Show_much}

Inspired by the best-fit rotation curve for FGC 227, we investigate the
possibility that all thick disks contain some fraction of
counter-rotating stars.  Our data is able to place tight constraints
on the amount of counter-rotating material since both the offplane
rotation curves and the midplane LOSVD will be strongly affected by
any counter-rotating stars.

In Section~\ref{Sstell_kin}, we imposed thin and thick disk flux
fractions based on previous photometric decompositions.  We now leave
the flux fractions as free parameters and instead hold the rotation
curve shapes fixed.  We fit two simple models, each with two
kinematically independent stellar components.  In the first model, we
assume there are two stellar components, one rotating identically as
the gas and one with zero net rotation, as one might expect for a
stellar halo.  The final observed rotation curve is a flux weighted
average of these two curves and we fit for the best fitting flux
ratio.  We restricted the explored parameter space such that the
rotation curves had to be some positive linear combination of the
midplane H$\alpha$ and a non-rotating or counter-rotating rotation
curve.  In the second model, we assume the second component is
counter-rotating with a velocity one-half the magnitude of the
H$\alpha$ rotation curve.  For both models, we calculate uncertainties
from the covariance matrix and scale them upwards such that the
reduced-$\chi^2$ equals unity (i.e., we assume our model should be a
good fit).  We do not calculate uncertainties when the fit converges
to a boundary condition.  We also do not construct detailed models for
cases like FGC 1415 where the stars could be better fit with a faster
rotation curve than the gas; these galaxies naturally converge on the
boundary condition of having no second component.  It should be
emphasized that these are simple toy models, and we have no direct
evidence of counter rotating thick disk stars beyond the strange
rotation curve of FGC 270.  For example, if we observed a MW like
(V$_{\rm{c}}=220$ \kms) galaxy that had a 10\% (by flux) thick disk
lagging at 40 \kms, we would compute a maximum counter-rotating
fraction of 1\% and a non-rotating fraction of 2\%, despite all the
stars being co-rotators.

\begin{figure*}
\epsscale{.65}
\plotone{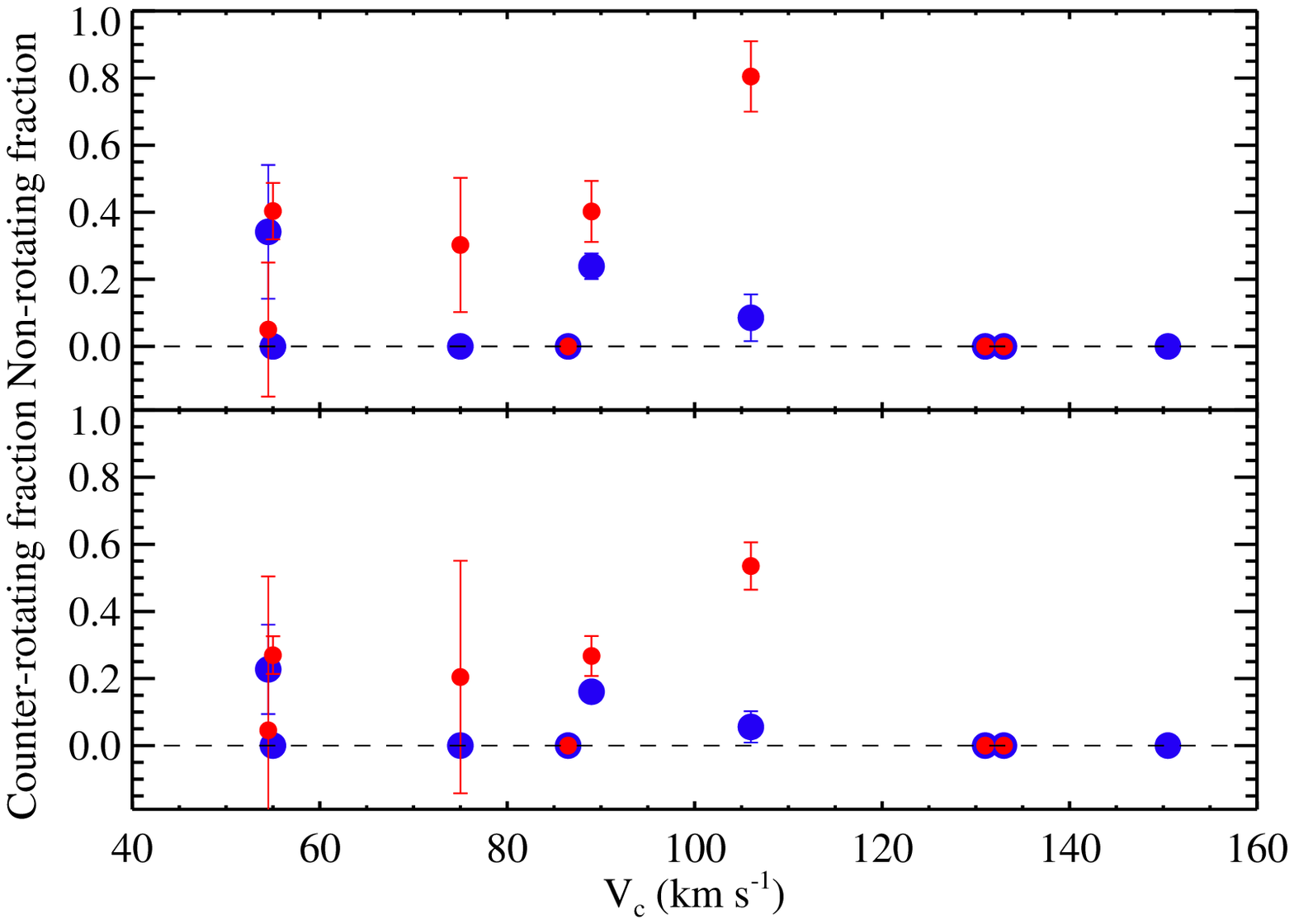}
\caption{Results from fitting the midplane and offplane rotation
curves as a combination of two fixed rotation curves.  In the top
panel, the rotation curves are a combination of the midplane H$\alpha$
and a flat non-rotating RC.  In the bottom panel, the base rotation
curves are the midplane H$\alpha$ combined with a rotation curve
counter-rotating with one-half the H$\alpha$ velocity.  These fits are
listed in Table~\ref{counter_table}. 
\label{counter_frac} }
\end{figure*}

The resulting fractions of non-rotating of counter rotating stars are
plotted in Figure~\ref{counter_frac} and are listed in
Table~\ref{counter_table}.  The midplane stellar rotation curves are
typically consistent with the H$\alpha$ rotation curve, with 6 of the
9 midplanes being best fit without a non-rotating or counter-rotating
component.  The remaining three galaxies do have midplane rotation
curves that are consistent with the presence of an additional lagging
component.  FGC 1948 is low mass with a surprisingly large LOSVD.  FGC
227 is the counter rotator with a LOSVD that dramatically increases
with radius.  FGC 2558 is the only galaxy to show a large discrepancy
between midplane and offplane H$\alpha$ rotation curves, has a stellar
lag that appears to be only on the receding side of the galaxy.

The offplane spectra show larger evidence for non- or counter-rotating
motion, with only 3 of the 9 galaxies requiring no slow rotating
component.  This effect can be seen in Figure~\ref{counter_frac},
where all of the offplane spectra show a preference for equal or
larger value of the counter-rotating fraction than seen in the
midplane.
%\clearpage

%\clearpage

\begin{deluxetable}{ c c c c c}
\tabletypesize{\scriptsize}
%\rotate
\tablewidth{0pt}
%\tablenum{num}
%\tablecolumns{num}
%\tableheadfrac{num}
\tablecaption{Non-Rotating and Counter Rotating Fractions \label{counter_table}}
\tablehead{\colhead{FGC} & \multicolumn{2}{c}{Non-Rotating Fraction}  &\multicolumn{2}{c}{Counter-Rotating Fraction}\\
            &\colhead{Thin Disk} & \colhead{Thick Disk} & \colhead{Thin Disk}&\colhead{Thick Disk}  }
\startdata
227 &  0.1$\pm$ 0.1 &  0.8$\pm$ 0.1 &  0.1$\pm$ 0.0 &  0.5$\pm$ 0.1  \\
780 &  0.0 &  0.3$\pm$ 0.2 &  0.0 &  0.2$\pm$ 0.3  \\
1415 &  0.0 &  0.0 &  0.0 &  0.0  \\
1440 &  0.0 &  \nodata & 0.0  &  \nodata  \\
1642 &  0.0 &  0.4$\pm$ 0.1 &  0.0 &  0.3$\pm$ 0.1  \\
1948 &  0.3$\pm$ 0.2 &  0.1$\pm$ 0.2 &  0.2$\pm$ 0.1 &  0.0$\pm$ 0.5  \\
2558 &  0.2$\pm$ 0.0 &  0.4$\pm$ 0.1 &  0.2$\pm$ 0.0 &  0.3$\pm$ 0.1  \\
E1371 &  0.0 &  0.0 &  0.0 &  0.0  \\
E1498 &  0.0 &  0.0 &  0.0 &  0.0  \\
\enddata
\end{deluxetable}
%\clearpage

\section{Expected Lags}\label{ex_lags}

Having found a wide range of thick disk behaviors, we now investigate the
expected stellar lags we should see in our sample of thick disks using a
dynamical model originally designed for the Milky Way.  The
large scale height of thick disk stars implies they have larger
velocity dispersions than thin disk stars.  If the larger vertical
velocity dispersion also reflects a larger radial velocity dispersion,
then the larger random motions of thick disk stars should lead to
their requiring less rotational support.  The thick disk stars should
therefore lag in velocity compared to the kinematically colder thin
disk stars and ionized gas.  

\citet{Girard06} use the Jeans equation and a series of reasonable
assumptions to model the expected thick disk lag in the MW as a
function of height above the midplane.  While this model was built to
explain the observed lag of thick disk stars in the Milky Way, the
formalism is easily generalizable to the galaxies in our sample.

Using the Jean's equation, \citet{Girard06} find that the rotational
velocity of a thick disk rotating in a Plummer dark matter potential
with an embedded thin disk is given by:
\begin{eqnarray}\label{jeaneq}
{\overline {v_{\Theta}}}^2 (z,R) = \sigma_R^2 (z)
\Big[
-{ \Upsilon_{a,b} R } \\+
0.5 \lambda \Big( 1 - {z \over h_{z_{thick}}} \Big) +
1 - {\sigma_{\Theta}^2 \over \sigma_R^2 }
\Big]  \nonumber \\
+ { (v_c^2 - v^2_{disk}(R,0))(R^2 + a^2)^{3/2}
\over (R^2 + z^2 +a^2)^{3/2} } +
v^2_{disk}(R,z) \nonumber ,
\end{eqnarray}
where $R$, $z$, and $\Theta$ are galactocentric cylindrical
coordinates.  The term ${\overline {v_{\Theta}}}$ is the average thick
disk velocity in the direction of galactic rotation, $\sigma_R$ and
$\sigma_\Theta$ are the radial and tangential components of velocity
dispersion for the thick disk stars, $v_c$ is the local standard of
rest velocity at the radius of interest, $v_{disk}$ is the portion of
the thick disk rotational velocity due to the gravitational potential
of the thin disk, $h_{z_{thick}}$ is the exponential thick disk scale
height and $a$ is the halo core radius.  The term $\Upsilon_{a,b}$
lets one approximate the thick disk as entirely self gravitating, or
gravitationally dominated by the embedded thin disk.  Because the
thick disk mass is small compared to the total gas and thin disk
mass in all of our galaxies, we choose to use $\Upsilon_{b}\sim2/h_R$.
The $\lambda$ term takes values of 1 or 0 in order to include or
exclude the velocity dispersion cross-term.

\begin{figure*}
\plotone{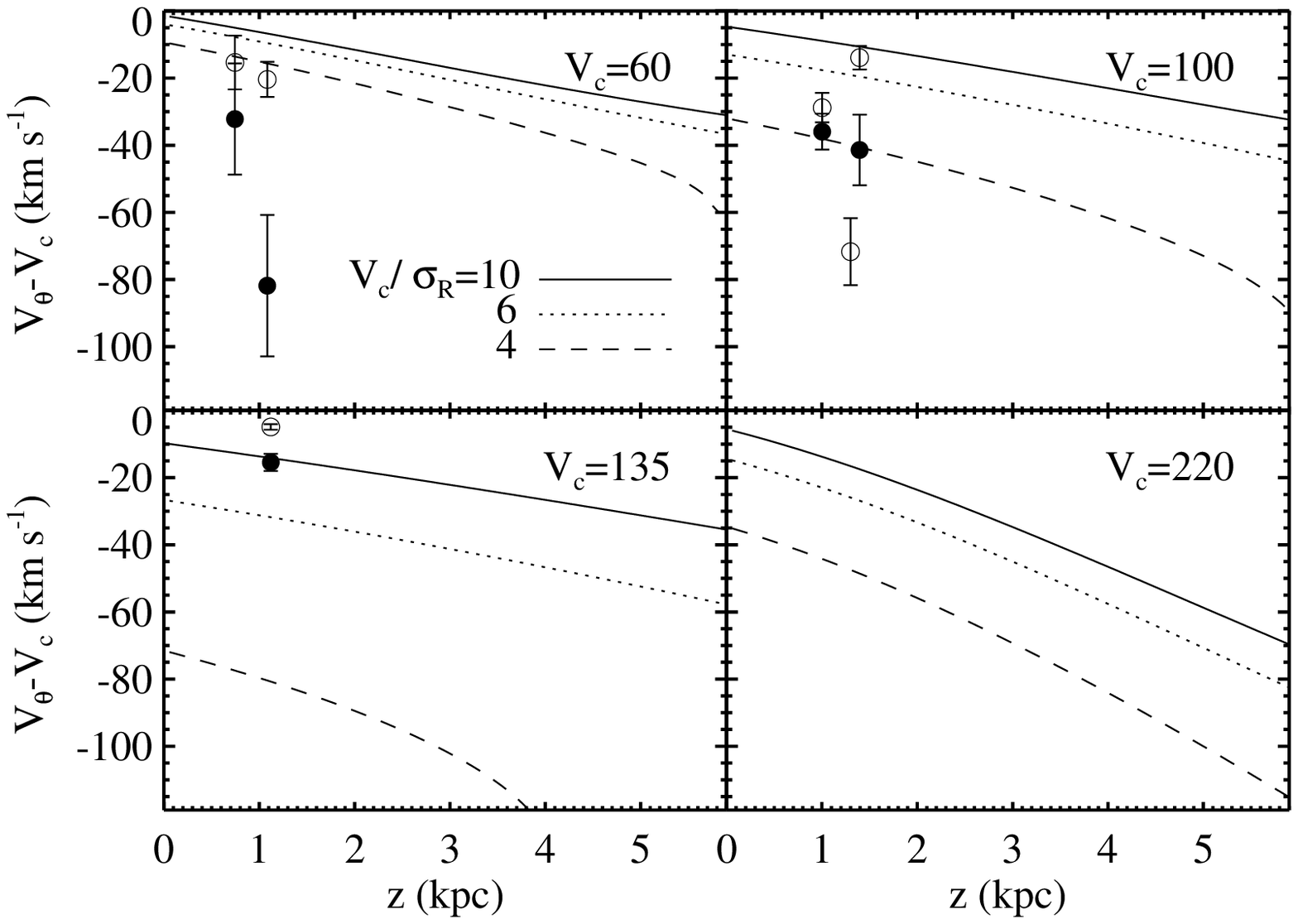}
\caption{The expected thick disk lags as a function of height above
the midplane and thick disk velocity dispersion.  The first three
panels show model galaxies similar to the ones in our sample.  Points
show the stellar lags measured from our rotation curve fits.  Open
points show lags from rotation curves where the offplane and midplane
rotation curves are fit independently.  Solid points show the average
lag for the models which correct for cross-contamination of the thin
and thick disk rotation curves, and are generally more reliable
estimates of the thick disk lag.  The final panel shows the results of
our model when we use MW like parameters.\label{gmod_fig} Observed
galaxies we compare to the models: In the upper left FGC 1642 and FGC
780; upper right FGC 1415, FGC 227, and FGC 2558; and lower left FGCE
1371.  All the models and observations are taken at $R=2.5 h_R$.  FGC
1948 is excluded from the plot because there is no coherent rotation.
FGC 1440 is excluded because we have no offplane stellar velocity
measurements.}
\end{figure*}

We calculate dynamical models for three fiducial galaxy masses and
three thick disk velocity dispersions.  We use realistic galactic
parameters taken from \citet{Yoachim06} to generate $h_{z_{thick}}$
and $v^2_{disk}(R,0)$.  For terms for which we do not have explicit
measurements, we use the approximation $a[\rm{kpc}]\approx
13(h_{r_{thin}}[\rm{kpc}]/5)^{1.05}$ given by \citet{Donato04}, assume
$\sigma_{\Theta}\sim\sigma_R$, and set $\lambda=0$.  We compute models
for different values of $\sigma_R$, as this is the dominant term in
producing stellar lags.  For simplicity, we assume the thick disk
velocity dispersion does not vary with height above the midplane.
This last approximation is not particularly valid given that
\citet{Girard06} find that the velocity dispersion in the MW increases
with a slope of 9 \kms kpc$^{-1}$.  However the difference between a
variable and constant velocity dispersion will be most pronounced at
large scale heights, beyond the range probed by our observations
($z\sim1.5-2$ kpc).  The resulting models are plotted in
Figure~\ref{gmod_fig} along with the lags we have measured in our
galaxies.  For reference, we also include a model using the same
assumptions but with morphology and velocities similar to the Milky
Way in Figure~\ref{gmod_fig}.

For most of the galaxies where we measure a thick disk lag,
Figure~\ref{gmod_fig} shows the thick disk kinematics could be well
explained by a population with radial velocity dispersion of between
15 and 30 \kms and $v_c/\sigma<4$.  As before, the major exception is
FGC 227.  The stellar lag for FGC 227 is so severe that it would imply
the thick disk is completely supported by random motions.  However, we
only detect flattened stellar populations in FGC 227, again consistent
with our interpretation that the thick disk is counter-rotating in
this system.

To verify that our model galaxies are reasonable, we use an identical
procedure to build a MW-like model.  Our MW-like model is a fair fit
to actual observations of the MW.  The measured thick disk velocity
dispersion in the solar neighborhood is 50 \kms, for which our model
correctly predicts the midplane thick disk lag of 30 \kms.  On the
other hand, the increase of the thick disk lag with scale height is
poorly fit by our model;  the observed lag increases with a slope of 30
\kms kpc$^{-1}$, and our model has a slope around half that.  This is
purely due to our choice to hold the velocity dispersion fixed--a
thick disk velocity dispersion that increased with height would
generate a more accurate slope.

%\clearpage
%\clearpage

Modeling the disks as cylindrically rotating is only a crude
approximation to account for the stellar cross-contamination.  In
reality, we expect the thin disk stars which reach large $z$ heights
to be the thin disk stars with larger velocity dispersions.  This
would mean that the thin disk stars at high $z$ should also be lagging
compared to the midplane thin disk stars.  Ideally, we would construct
a fully self-consistent dynamical model of each galaxy, but our large
uncertainties and limited LOSVD information would result in model
degeneracies.  Constructing a robust self consistent dynamical model
of a galaxy also benefits from larger numbers of data points
\citep{Girard06}.  With only a handful of stellar rotation curve
points per galaxy, we do not have enough data to constrain a more
complex model.  We simply point out that when we correct for the
cross-contamination of the rotation curves we may be over-correcting
the data.  We estimate the magnitude of the over-correction using
dynamical models in \S~\ref{ex_lags}.

\section{Dust and Projection Effects }\label{Sdust}

\begin{figure}
\plotone{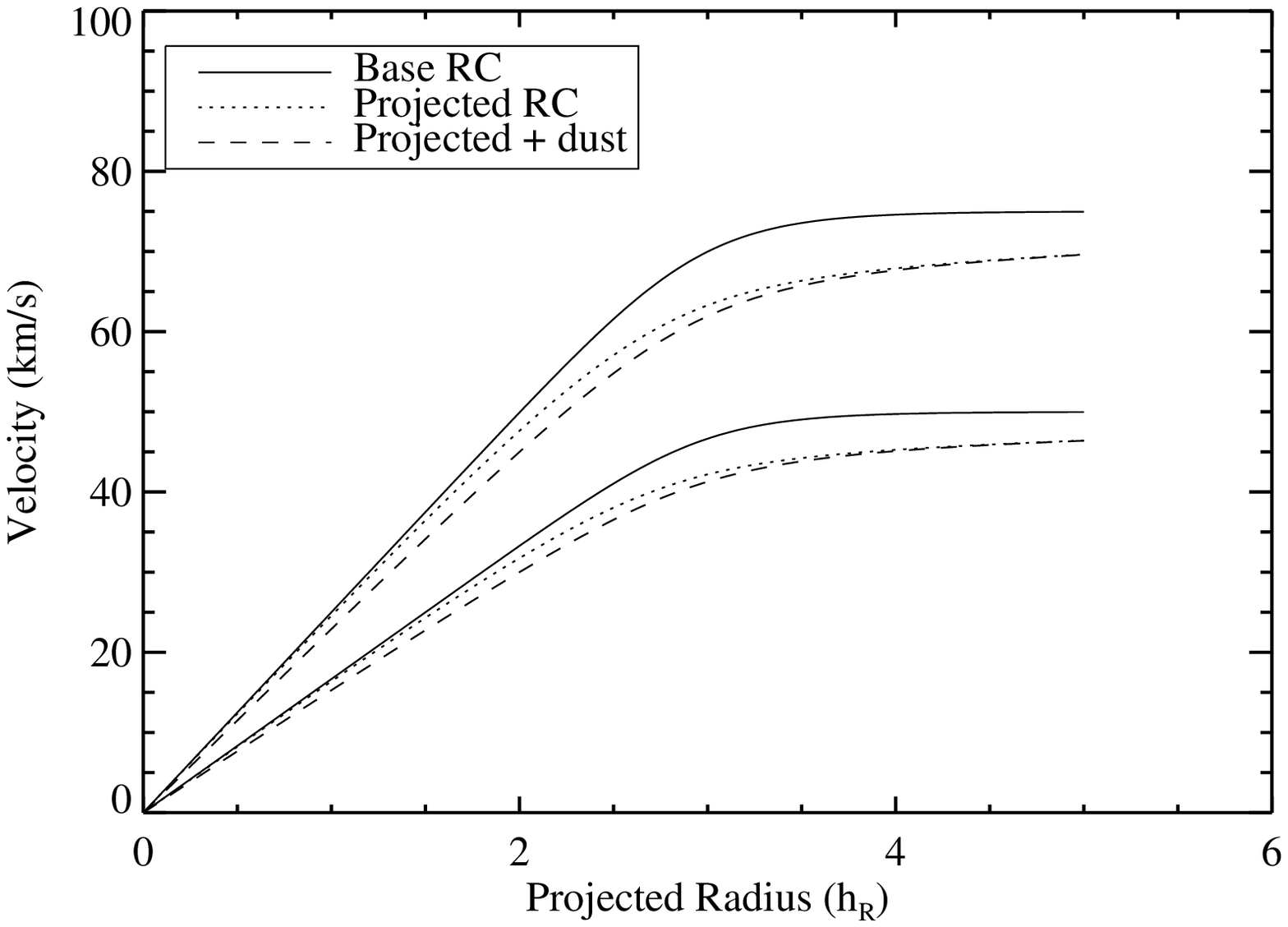}
\caption{Two examples of the effects dust and projection can have on
our observed rotation curves.  While projection creates considerable
changes, the addition of dust extinction is negligible.  \label{exrc}}
\end{figure}

As a final check that our observed kinematics indeed reflect the true
stellar motions, we now explore the expected impact of projection
effects and dust extinction, both of which can create differences
between the observed and underlying rotation curves.  In
Figure~\ref{exrc}, we show how two input rotation curves are modified
by being viewed edge-on, with and without dust.  For these models, we
assumed an exponential disk of stars and dust, and for simplicity only
considered absorption (i.e., ignoring scattering).  The amount of dust
adopted in the model would generate an extinction of 2.2 magnitudes in
the total apparent magnitude of the galaxy.  This is a rather large
extinction for the near-IR, given that the observed galaxies in our
sample are only offset by 0.2 mag from the face-on NIR Tully-Fisher
relation \citep{Yoachim06}.  We adopted an underlying rotation curve
shape from \citet{Courteau97}.  

As can be seen from Figure~\ref{exrc}, the inner regions of the
rotation curve are generally unchanged due to projection, and the only
significant changes happen in the outer regions, where the true rotation
curve is flat.  These projection effects create a lag of 7.2 \kms.

We have not corrected our rotation curves for these projection
effects, as we are primarily interested in the differences between the
thin and thick disk rotation curves.  This could lead us to make
systematic errors in interpreting the rotation curves if the
morphologies of the thin and thick disk are radically different, but
we have no reason to assume this is the case.

When dust is added to the model, it only creates an additional 2.6
\kms\ lag, in spite of the very high extinction adopted here.  This
model is completely consistent with the results of \citet{Matthews01},
who found that projection effects are dominant compared to extinction
in edge-on systems.  We do not expect our sample galaxies to have
larger extinctions than what is modeled in Figure~\ref{exrc}.

Full radiative transfer models \citep{Kregel05d, Bianchi07,
Xilouris99}, as well as comparisons of gaseous and optical rotation
curves \citep{Bosma92} have consistently found massive disk galaxies
have a central face-on optical depth near unity in the $V$-band, with
lower extinction levels in less massive systems like those that
dominate our sample \citep{Calzetti01}.  Dust levels this low should
not be expected to alter the observed rotation curve significantly,
even if a galaxy is viewed edge-on.  Moreover, most of our offplane
rotation curves exhibit a lag compared to the midplane.  In contrast,
If there were strong dustlanes affecting our midplane observations
(and not the offplane), the midplane would be the lagging component.

The combination of working at near-IR wavelengths, offsetting our slit
from any prominent dustlanes, and observing intrinsically linearly
rising rotation curves means our rotation curves should be fairly
unaffected by extinction or projection.  However, the same cannot be said
for our measured line-of-sight velocity dispersion (LOSVD).  Unlike
the rotational velocity measurement, which is mostly unaffected by
flux contributions from different radii, we expect the LOSVD to be
significantly broadened by projection effects.  We also find that in
most of our galaxies the LOSVD is very close to the instrumental
resolution, making any interpretation of the velocity dispersion
suspect.  Because of these challenges, we limit our analysis of the
LOSVD to only those cases where we believe our measurements are of
high quality and not dominated by the instrumental dispersion.

%\clearpage
%\clearpage

\section{Discussion}

The results of Sections~\ref{Srcs} and \ref{Sstell_kin} show that
stellar kinematics above the midplane display a wide range of
behaviors.  In higher mass systems (FGCE 1371, FGCE 1498, FGC 1440),
our midplane and offplane spectra show no clear signature of a hot
thick disk component.  The stellar rotation curves for these galaxies
are well matched by the midplane ionized gas H$\alpha$ RCs at all
measured scale heights.  All three of these galaxies converged to
models where the rotation curves contain no lagging component
(Table~\ref{counter_table}). However, \citet{Yoachim06} found that the
stellar flux in higher mass galaxies are dominated by the thin disk
component.  Therefore, the lack of a significant lag in these systems
is likely a result of the kinematically cold thin disk dominating the
stellar flux to scale heights of 1 kpc.  This result is not completely
unexpected, as the MW thin and thick disks should have similar
luminosities 1 kpc off the midplane \citep{Juric08}. We note that
there is still ample photometric evidence that these higher mass
galaxies contain thick disks, but they are simply too faint relative
to the thin disk for modest kinematic lags to be detected
spectroscopically.

The low mass galaxies in our sample do show measurable differences
between the midplane and offplane observations.  At large radii, we
find several galaxies where the offplane component is lagging compared
to the midplane (Figure~\ref{gmod_fig}).  In three of the low mass
systems (FGC 1415, FGC 1642, and FGC 780), the lags in the offplane
observations become more pronounced when we correct for the expected
thin disk contamination.  These lags are consistent with those that
are expected from dynamics alone (Equation~\ref{jeaneq}), provided
that the thick disk has a radial velocity dispersion between 15 and 30
\kms (i.e., 10-25\% of $v_c$).  Thus, the lags in these systems do not
necessarily require the presence of any counter-rotating material,
although a small amount of such material could be present.  FGC 2558
may also fall into this category; however the offplane RC is very
similar to the midplane, implying this could be another galaxy where
we have not successfully isolated the thick disk.  The observed lags
were easier to detect in these lower mass systems, due to their more
prominent thick disks.

The final two low mass galaxies in our sample, FGC 227 and FGC 1948,
have remarkably different rotation curves between the midplane and
offplane.  FGC 1948 does not display coherent stellar rotation in
either the midplane or the offplane, and therefore our subsequent fits
converge to extreme, and probably incorrect, models.  FGC 227 does
show rotation on the midplane, and a very low level of net rotation on
the offplane.  Our best fitting model for this galaxy has the thick
disk counter-rotating relative to the thin disk, consistent with the
radially increasing LOSVD which is a signature of unresolved counter
rotating stellar components.

Our measurements of the LOSVD are less than enlightening.  With the
exception of the radial increase in the LOSVD in FGC 227 and the high
LOSVD in FGC 1948, the rest of our LOSVD measurements show no
significant trends with radius and are close to the instrumental
resolution limit, suggesting that the radial velocity dispersions of
both the thin and thick disks are cold enough that we cannot reliably
measure their velocity dispersions at our spectral resolution.

Given the above results, our galaxies can be described as falling into
three categories: The high mass systems which have little to no thick
disk lag (or, more likely, thick disks which are so faint that we have
failed to measure their kinematics); the moderately lagging systems; and the
counter rotating system.  We can now compare these
results to the predictions of popular formation models for the
thick disk.

If thick disks are the result of gradual stochastic heating, we would
expect to always find thick disks co-rotating with the embedded thin
disks.  Moreover, with stronger spiral arms, larger molecular clouds,
and more massive dark matter substructure, the high mass systems
should be able to efficiently heat their thin disk stars into a
thicker disk.  Instead, we have found the opposite, with more
prominent thick disks and larger lags in the lower mass systems, as
well as evidence for counter-rotating stars.  This seems to rule out
gradual heating as the dominant method of thick disk formation,
particularly for low mass galaxies.

Forming thick disks in major mergers also does a poor job explaining
our observations. If thick disks were predominantly formed in major
mergers that disrupt and heat previously thin disks, we should expect
to find galaxies that never formed a thick disk, or that have failed
to accrete and cool enough gas to rebuild their thin disk components.
Major mergers also typically result in the formation of centrally
concentrated spheroidal components, making them a poor mechanism for
forming thick disks in the bulgeless galaxies observed here.

Unlike the two heating models, the variety of thick disk kinematics is
compatible with minor mergers and/or accretion.  Presumably, the thick
disk kinematics we observe are simply the kinematics left over from
the accretion event which deposited the majority of thick disk stars
or which triggered the formation of stars from gas accreted at large
scale heights.  The wide variety of possible accretion events
(co-rotating vs counter rotating, early disruption vs late disruption,
high eccentricity vs circular initial orbit) can evolve into
virialized thick disks with kinematics that are sometimes decoupled
from the thin disks and that show large variation from galaxy to
galaxy.  The ubiquity of thick disks is also well explained by the
merger/accretion scenario, given that galaxy formation in a $\Lambda$CDM
cosmology is dominated by hierarchical merging, and predicts that
every galaxy has a rich merger history.

Although the available data all points to a merger/accretion origin
for the thick disk, it is difficult to disentangle models where thick
disk stars are directly accreted from those where the stars form
{\emph{in situ}} further off the midplane during gas rich mergers
\citep{Brook04}.  

This ambiguity results from two sources.  First, there is no clear
dividing line between what one calls a star-forming region off the
midplane and a merging star-forming satellite galaxy.  Second, we know
from \citet{Yoachim06} that at least 75-90\% of the baryonic accretion
onto the galaxies was gaseous, and some fraction of this was certainly
accreted in bound subhalos.  Stars that formed initially in subhalos
before being accreted are likely to have kinematics similar to those
that formed from accreted gas during those same merging events.
Presumably, one could use detailed stellar age and abundance
information to help, but unfortunately this is only possible for the
closest galaxies.

There is evidence that much of the brighter inner halo and outer disk
substructure of M31 was formed through accretion \citep{Ferguson02,Koch07}.
These features would probably resemble a thick disk if M31 were more
distant and the features were unresolved.  Taking this lesson from
nearby galaxies, it is clear we are using smooth functions to
describe thick disk that may actually be highly structured systems.
However, the smooth descriptions of thick disks still provide a
reasonable statistical description of the ensemble of accreted stars.

In this study, we have measured thick disk kinematics in only very
late-type disk systems.  However, thick disks have been
photometrically detected in a wide variety of Hubble types
\citep[e.g.,][]{Seth05b,Pohlen04,Morrison97,vanDokkum94}.  The
kinematics in our sample are most consistent with merger/accretion
forming for the thick disks, but, except for the Milky Way, there have
been no measurements of thick disk kinematics in earlier type
galaxies.

By focusing on disk systems, we may not be sensitive to how thick
disks form across all Hubble types.  Almost by definition, late-type
galaxies have not suffered a major-merger since the formation of their
stellar disks, otherwise they would likely possess large spheroidal
components and be classified as an earlier type system.  The only way
pure disk galaxies could form thick disks is either through accretion
or stochastic heating.

\section{Conclusions}

We have expanded the kinematic observations of \citet{Yoachim05} to
include a total of nine galaxies with thick disks.  Analyzing our low
signal-to-noise spectra that contain systematic sky line
residuals prompted us to develop a brute-force method of
cross-correlation to extract stellar rotation curves.  In
galaxies with $V_c> 120$ km s$^{-1}$, we do not detect any measurable
difference between the thin and thick disk stellar kinematics.  This
is most likely due to a combination of thin disks being brighter in
more massive galaxies, and the expected change in rotation curve as a
function of scale height being smaller.

In lower mass galaxies ($V_c< 120$ km s$^{-1}$), we find a variety of
thick disk behaviors.  Thick disks are found with both small and
large magnitude lags, including a counter-rotating thick disk.

The observed kinematics are best explained by thick disk formation
models where the thick disks in low mass systems are composed of stars
that have been accreted from satellite galaxies or are formed at large
scale heights from accreting gas.  Models where the thick disks form
during major mergers or through stochastic heating seems unable to
explain the wide range of thick disk kinematics we observe.  While we
strongly favor a formation model of thick disks via accretion, we
stress that this result can not necessarily be generalized to other
Hubble types or higher mass systems ($V_c> 120$ km s$^{-1}$).

%--------------------------------------
\acknowledgments

We thank the Gemini support staff for their help preparing and
executing these observations.  We thank Suzanne Hawley for reading an
early version of this paper and making helpful comments.  We also
thank the anonymous referee for helpful comments.  JJD and PY were
partially supported through NSF grant CAREER AST-0238683 and the
Alfred P.\ Sloan Foundation.  Based on observations obtained at the
Gemini Observatory, which is operated by the Association of
Universities for Research in Astronomy, Inc., under a cooperative
agreement with the NSF on behalf of the Gemini partnership: the
National Science Foundation (United States), the Particle Physics and
Astronomy Research Council (United Kingdom), the National Research
Council (Canada), CONICYT (Chile), the Australian Research Council
(Australia), CNPq (Brazil), and CONICET (Argentina).  This research
used the facilities of the Canadian Astronomy Data Centre operated by
the National Research Council of Canada with the support of the
Canadian Space Agency.

\appendix 

\section{Stellar Rotation Curves in the Presence of Systematic 
Errors }\label{ap1}

Working in the near-IR, we find our spectra have regions which are
dominated by both Gaussian and systematic errors caused by bright
atmospheric emission lines.  To properly measure stellar kinematics
based on spectral absorption features we must employ a method that is
not affected by our sky line residuals.

There are two common techniques of deriving the kinematic information
from galaxy spectra--direct $\chi^2$-fitting and cross-correlation.
In direct $\chi^2$-fitting \citep{Rix92,Kelson00,Barth02,
Cappellari04}, a template star is redshifted and broadened to fit a
galaxy spectrum, while in cross-correlation techniques
\citep{Simkin74,Tonry79,Statler95} a template star is cross-correlated
with the galaxy spectrum and the kinematic properties are deduced from
the position and shape of the cross-correlation peak.

Cross-correlation techniques have the advantage of being
computationally efficient, often making use of fast Fourier transform
algorithms.  The cross-correlation technique benefits greatly from the
fact that the Fourier transform of Gaussian noise is also Gaussian
noise.  In this way, noise in the galaxy spectrum transforms into
random noise in the cross-correlation while the kinematic information
becomes concentrated in a central peak.  However, this is only true if the
noise is uniform throughout the spectrum.  Using a direct chi-squared
fit is more computationally expensive, but has the added benefit of
being able to weight individual wavelengths according to their
specific signal-to-noise, or completely mask wavelengths that are
affected by systematic errors.

Although direct chi-squared fitting works well in some situations, at
low S/N($<$20), any direct chi-squared fitting routine will
over-smooth the data because the low S/N continuum is best fit by a
strait line (i.e., an over-broadened template star).  In previous
studies that have used direct fitting, \citet{Kelson00} has a median
S/N of 35/\AA, while \citet{Barth02} report a S/N/pixel of 100-200.
In contrast, our data has SNR$<20$/\AA, due to the very low surface
brightness of our targets.

Because we have both low S/N and regions which require masking, we
have created a fitting procedure which utilizes cross-correlation
without making use of the computational time saving FFT techniques of
previous authors.

Traditional cross-correlation of discrete functions is defined as
\begin{equation}
(f\star g)_i \equiv \sum_j f_j g_{i+j}.
\end{equation}

We adopt a normalized version, where the means of the spectra have
been subtracted before the cross-correlation is computed
\begin{equation}
(f\star g)_L=\frac{\sum_{k=1}^{N-L} f_{L} g_{k+L}}
{\sqrt{\sum_{k=1}^{N}(f_k)^2
\sum_{k=1}^{N}(g_k)^2 } },
\end{equation}
where N is the number of points in the given spectra.  For lags less
than zero, the numerator becomes $\sum_{k=1}^{N-|L|} f_{k+|L|} g_{k}$.
This ensures spectra with perfectly matching shapes will have a
maximum cross-correlation amplitude of unity.

%\clearpage
\begin{figure*}
\epsscale{.5}
\plotone{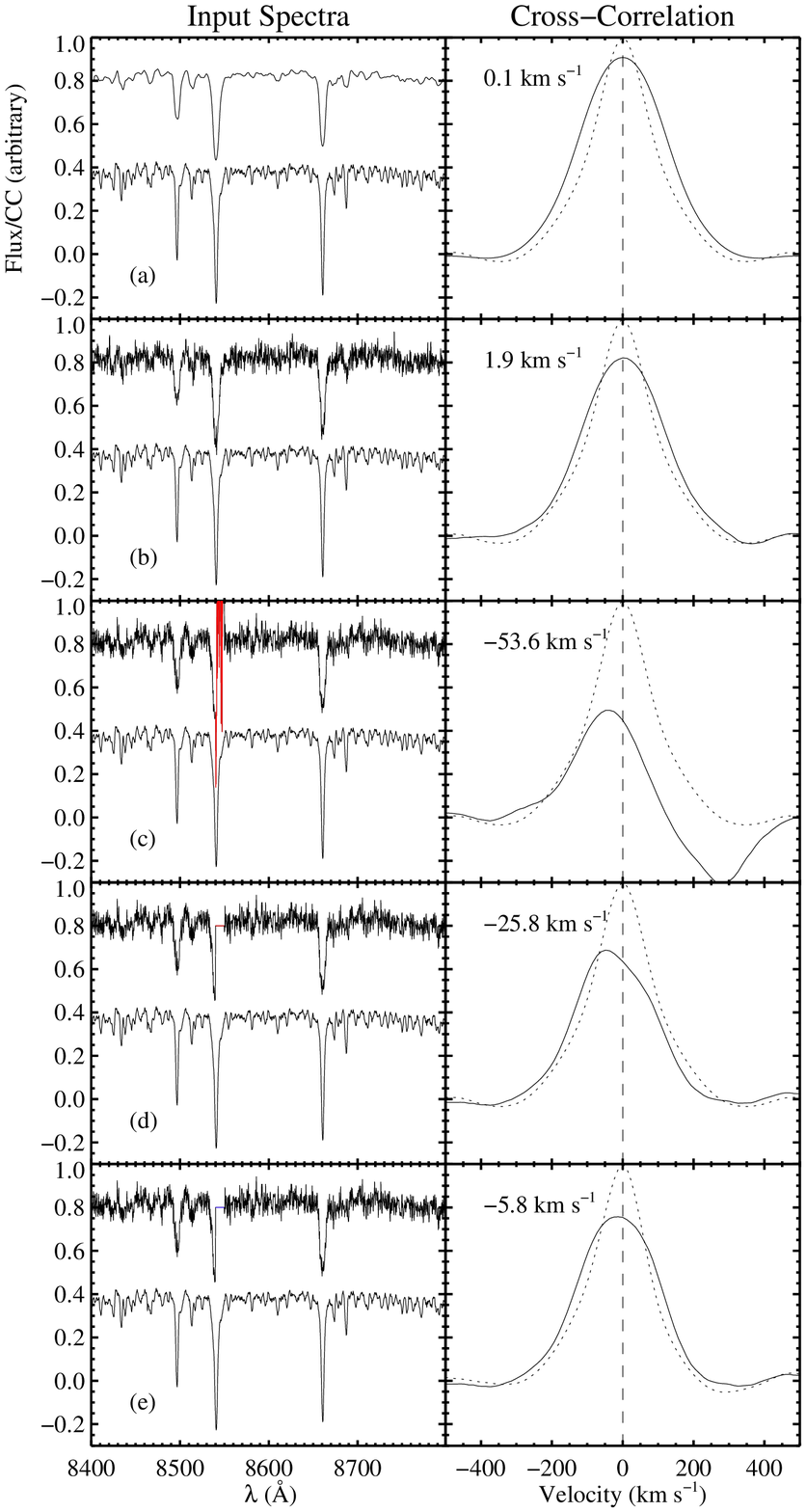}
\caption{Examples of cross-correlating in the presence of different
types of noise.  In the left hand column, we show a model galaxy
spectrum (top) and stellar template (bottom).  In the right hand
column, we plot the galaxy-star cross-correlation (solid) and stellar
auto-correlation (dotted) and note the velocity error resulting from
comparing the two.  (a) The ideal case of a high signal-to-noise
galaxy spectrum.  (b) Results from a galaxy spectrum with a
S/N/\AA$\sim10$.  (c) Spectra with a small region of very low
S/N affecting a section of one of the Ca absorption features, similar
to how bright sky lines leave residuals on our spectra.  (d) A
traditional cross-correlation where the noisy region has been set to
the continuum.  (e) Our new cross-correlation technique where we
compute the cross-correlation excluding the masked
region.  \label{cc_example}}
\end{figure*}
%\clearpage

Finally, we define masks $\delta$ for each spectrum which have values
of 1 in regions of good data and 0 for masked wavelengths.  Given a
stellar spectrum $S$ and Galaxy spectrum $G$ that are binned in logarithmic
wavelength intervals and have both been normalized by division of a
low order polynomial and had their means subtracted, we compute our
modified cross-correlation as
\begin{equation}
(S\star G)_L=\frac{\sum_{k=1}^{N-L} S_{L} G_{k+L} \delta^S_L \delta^G_{K+L}}
{\sqrt{\sum_{k=1}^{N}(S_k \delta^S_K \delta^G_K)^2
\sum_{k=1}^{N}(G_k \delta^S_K \delta^G_K)^2 } }.
\end{equation}

We then generate a model galaxy spectrum $M$ by redshifting and
broadening the stellar template, $M(x)=S(x+v)\otimes B(x)$ where $B(x)$
is a Gaussian broadening function, $v$ is a velocity shift, and
$\otimes$ represents convolution.  We then calculate the model's
modified cross-correlation using the masks from the actual galaxy
spectrum
\begin{equation}
(S\star M)_L=\frac{\sum_{k=1}^{N-L} S_{L} M_{k+L} \delta^S_L \delta^G_{K+L}}
{\sqrt{\sum_{k=1}^{N}(S_k \delta^S_K \delta^G_K)^2
\sum_{k=1}^{N}(M_k \delta^S_K \delta^G_K)^2 } }.
\end{equation}

We vary the velocity shift and broadening to minimize the $\chi^2$
between $(S\star G)$ and $(S\star M)$.  We focus on the region of
the primary peak, and clip regions beyond the bracketing local minima.
Examples of traditional cross-correlation and our modified
cross-correlation are shown in Figure~\ref{cc_example}.  In general,
our masked cross-correlation technique cannot reproduce the excellent
fits that are possible with data that is unaffected by systematics,
but we can reduce the errors to be of order 5 \kms\ in our typical
spectra.

%bibtex code:
%\bibliography{/home/prawn/yoachim/Papers/Bib_files/big_jabref}

\begin{thebibliography}{95}
\expandafter\ifx\csname natexlab\endcsname\relax\def\natexlab#1{#1}\fi

\bibitem[{Abadi {et~al.}(2003)Abadi, Navarro, Steinmetz, \& Eke}]{Abadi203}
Abadi, M.~G., Navarro, J.~F., Steinmetz, M., \& Eke, V.~R. 2003, \apj, 597, 21

\bibitem[{Abe {et~al.}(1999)Abe, Bond, Carter, Dodd, Fujimoto, Hearnshaw,
  Honda, Jugaku, Kabe, Kilmartin, Koribalski, Kobayashi, Masuda, Matsubara,
  Miyamoto, Muraki, Nakamura, Nankivell, Noda, Pennycook, Pipe, Rattenbury,
  Reid, Rumsey, Saito, Sato, Sato, Sekiguchi, Sullivan, Sumi, Watase,
  Yanagisawa, Yock, \& Yoshizawa}]{Abe99}
Abe, F., Bond, I.~A., Carter, B.~S., Dodd, R.~J., Fujimoto, M., Hearnshaw,
  J.~B., Honda, M., Jugaku, J., Kabe, S., Kilmartin, P.~M., Koribalski, B.~S.,
  Kobayashi, M., Masuda, K., Matsubara, Y., Miyamoto, M., Muraki, Y., Nakamura,
  T., Nankivell, G.~R., Noda, S., Pennycook, G.~S., Pipe, L.~Z., Rattenbury,
  N.~J., Reid, M., Rumsey, N.~J., Saito, T., Sato, H., Sato, S., Sekiguchi, M.,
  Sullivan, D.~J., Sumi, T., Watase, Y., Yanagisawa, T., Yock, P.~C.~M., \&
  Yoshizawa, M. 1999, \aj, 118, 261

\bibitem[{Barbieri {et~al.}(2005)Barbieri, Fraternali, Oosterloo, Bertin,
  Boomsma, \& Sancisi}]{Barbieri05}
Barbieri, C.~V., Fraternali, F., Oosterloo, T., Bertin, G., Boomsma, R., \&
  Sancisi, R. 2005, \aap, 439, 947

\bibitem[{Barth {et~al.}(2002)Barth, Ho, \& Sargent}]{Barth02}
Barth, A.~J., Ho, L.~C., \& Sargent, W.~L.~W. 2002, \aj, 124, 2607

\bibitem[{Bekki \& Chiba(2001)}]{Bekki01}
Bekki, K., \& Chiba, M. 2001, \apj, 558, 666

\bibitem[{Bensby {et~al.}(2003)Bensby, Feltzing, \& Lundstr\"om}]{Bensby03}
Bensby, T., Feltzing, S., \& Lundstr\"om, I. 2003, \aap, 410, 527

\bibitem[{Bensby {et~al.}(2005)Bensby, Feltzing, Lundstr\"om, \&
  Ilyin}]{Bensby05}
Bensby, T., Feltzing, S., Lundstr\"om, I., \& Ilyin, I. 2005, \aap, 433, 185

\bibitem[{Benson {et~al.}(2004)Benson, Lacey, Frenk, Baugh, \& Cole}]{Benson04}
Benson, A.~J., Lacey, C.~G., Frenk, C.~S., Baugh, C.~M., \& Cole, S. 2004,
  \mnras, 351, 1215

\bibitem[{{Bianchi}(2007)}]{Bianchi07}
{Bianchi}, S. 2007, \aap, 471, 765

\bibitem[{Bosma {et~al.}(1992)Bosma, Byun, Freeman, \& Athanassoula}]{Bosma92}
Bosma, A., Byun, Y., Freeman, K.~C., \& Athanassoula, E. 1992, \apjl, 400, L21

\bibitem[{Brewer \& Carney(2004)}]{Brewer04}
Brewer, M., \& Carney, B.~W. 2004, Publications of the Astronomical Society of
  Australia, 21, 134

\bibitem[{Brewer \& Carney(2006)}]{Brewer06}
Brewer, M.-M., \& Carney, B.~W. 2006, \aj, 131, 431

\bibitem[{Brook {et~al.}(2004)Brook, Kawata, Gibson, \& Freeman}]{Brook04}
Brook, C.~B., Kawata, D., Gibson, B.~K., \& Freeman, K.~C. 2004, \apj, 612, 894

\bibitem[{Burstein(1979)}]{Burstein79}
Burstein, D. 1979, \apj, 234, 829

\bibitem[{{Calzetti}(2001)}]{Calzetti01}
{Calzetti}, D. 2001, \pasp, 113, 1449

\bibitem[{Cappellari \& Emsellem(2004)}]{Cappellari04}
Cappellari, M., \& Emsellem, E. 2004, \pasp, 116, 138

\bibitem[{{Carlberg}(1987)}]{Carlberg87}
{Carlberg}, R.~G. 1987, \apj, 322, 59

\bibitem[{Cescutti {et~al.}(2007)Cescutti, Matteucci, Fran\c~cois, \&
  Chiappini}]{Cescutti07}
Cescutti, G., Matteucci, F., Fran\c~cois, P., \& Chiappini, C. 2007, \aap, 462,
  943

\bibitem[{Chen {et~al.}(2001)Chen, Stoughton, Smith, Uomoto, Pier, Yanny,
  Ivezi\'c, York, Anderson, Annis, Brinkmann, Csabai, Fukugita, Hindsley,
  Lupton, Munn, \& the SDSS~Collaboration}]{Chen01}
Chen, B., Stoughton, C., Smith, J.~A., Uomoto, A., Pier, J.~R., Yanny, B.,
  Ivezi\'c, v.~Z., York, D.~G., Anderson, J.~E., Annis, J., Brinkmann, J.,
  Csabai, I., Fukugita, M., Hindsley, R., Lupton, R., Munn, J.~A., \& the
  SDSS~Collaboration. 2001, \apj, 553, 184

\bibitem[{Chiappini {et~al.}(1997)Chiappini, Matteucci, \&
  Gratton}]{Chiappini97}
Chiappini, C., Matteucci, F., \& Gratton, R. 1997, \apj, 477, 765

\bibitem[{Chiba \& Beers(2000)}]{Chiba00}
Chiba, M., \& Beers, T.~C. 2000, \aj, 119, 2843

\bibitem[{Courteau(1997)}]{Courteau97}
Courteau, S. 1997, \aj, 114, 2402

\bibitem[{Dalcanton \& Bernstein(2000)}]{Dalcanton00}
Dalcanton, J.~J., \& Bernstein, R.~A. 2000, \aj, 120, 203

\bibitem[{Dalcanton \& Bernstein(2002)}]{Dalcanton02}
---. 2002, \aj, 124, 1328

\bibitem[{Dalcanton {et~al.}(2004)Dalcanton, Yoachim, \&
  Bernstein}]{Dalcanton04}
Dalcanton, J.~J., Yoachim, P., \& Bernstein, R.~A. 2004, \apj, 608, 189

\bibitem[{de~Grijs \& Peletier(1997)}]{deGrijs97b}
de~Grijs, R., \& Peletier, R.~F. 1997, \aap, 320, L21

\bibitem[{de~Grijs \& van~der Kruit(1996)}]{deGrijs96}
de~Grijs, R., \& van~der Kruit, P.~C. 1996, \aaps, 117, 19

\bibitem[{Donato {et~al.}(2004)Donato, Gentile, \& Salucci}]{Donato04}
Donato, F., Gentile, G., \& Salucci, P. 2004, \mnras, 353, L17

\bibitem[{Elmegreen \& Elmegreen(2006)}]{Elmegreen06}
Elmegreen, B.~G., \& Elmegreen, D.~M. 2006, \apj, 650, 644

\bibitem[{Fall \& Efstathiou(1980)}]{Fall80}
Fall, S.~M., \& Efstathiou, G. 1980, \mnras, 193, 189

\bibitem[{Feltzing {et~al.}(2003)Feltzing, Bensby, \& Lundstr\"om}]{Feltzing03}
Feltzing, S., Bensby, T., \& Lundstr\"om, I. 2003, \aap, 397, L1

\bibitem[{{Ferguson} {et~al.}(2002){Ferguson}, {Irwin}, {Ibata}, {Lewis}, \&
  {Tanvir}}]{Ferguson02}
{Ferguson}, A.~M.~N., {Irwin}, M.~J., {Ibata}, R.~A., {Lewis}, G.~F., \&
  {Tanvir}, N.~R. 2002, \aj, 124, 1452

\bibitem[{Fraternali \& Binney(2006)}]{Fraternali06}
Fraternali, F., \& Binney, J.~J. 2006, \mnras, 366, 449

\bibitem[{Freeman \& Bland-Hawthorn(2002)}]{Freeman02}
Freeman, K., \& Bland-Hawthorn, J. 2002, \araa, 40, 487

\bibitem[{Geha {et~al.}(2005)Geha, Guhathakurta, \& van~der Marel}]{Geha05}
Geha, M., Guhathakurta, P., \& van~der Marel, R.~P. 2005, \aj, 129, 2617

\bibitem[{Gilmore \& Reid(1983)}]{Gilmore83}
Gilmore, G., \& Reid, N. 1983, \mnras, 202, 1025

\bibitem[{Gilmore {et~al.}(2002)Gilmore, Wyse, \& Norris}]{Gilmore02}
Gilmore, G., Wyse, R.~F.~G., \& Norris, J.~E. 2002, \apjl, 574, L39

\bibitem[{Girard {et~al.}(2006)Girard, Korchagin, Casetti-Dinescu, van Altena,
  L\'opez, \& Monet}]{Girard06}
Girard, T.~M., Korchagin, V.~I., Casetti-Dinescu, D.~I., van Altena, W.~F.,
  L\'opez, C.~E., \& Monet, D.~G. 2006, \aj, 132, 1768

\bibitem[{Glazebrook \& Bland-Hawthorn(2001)}]{Glazebrook01}
Glazebrook, K., \& Bland-Hawthorn, J. 2001, \pasp, 113, 197

\bibitem[{H{\"a}nninen \& Flynn(2002)}]{Hann02}
H{\"a}nninen, J., \& Flynn, C. 2002, \mnras, 337, 731

\bibitem[{{Hayashi} \& {Chiba}(2006)}]{Hayashi06}
{Hayashi}, H., \& {Chiba}, M. 2006, \pasj, 58, 835

\bibitem[{Heald {et~al.}(2006{\natexlab{a}})Heald, Rand, Benjamin, \&
  Bershady}]{Heald06b}
Heald, G.~H., Rand, R.~J., Benjamin, R.~A., \& Bershady, M.~A.
  2006{\natexlab{a}}, \apj, 647, 1018

\bibitem[{{Heald} {et~al.}(2007){Heald}, {Rand}, {Benjamin}, \&
  {Bershady}}]{Heald07}
{Heald}, G.~H., {Rand}, R.~J., {Benjamin}, R.~A., \& {Bershady}, M.~A. 2007,
  \apj, 663, 933

\bibitem[{Heald {et~al.}(2006{\natexlab{b}})Heald, Rand, Benjamin, Collins, \&
  Bland-Hawthorn}]{Heald06}
Heald, G.~H., Rand, R.~J., Benjamin, R.~A., Collins, J.~A., \& Bland-Hawthorn,
  J. 2006{\natexlab{b}}, \apj, 636, 181

\bibitem[{{Ibata} {et~al.}(2005){Ibata}, {Chapman}, {Ferguson}, {Lewis},
  {Irwin}, \& {Tanvir}}]{Ibata05}
{Ibata}, R., {Chapman}, S., {Ferguson}, A.~M.~N., {Lewis}, G., {Irwin}, M., \&
  {Tanvir}, N. 2005, \apj, 634, 287

\bibitem[{{Juri{\'c}} {et~al.}(2008){Juri{\'c}}, {Ivezi{\'c}}, {Brooks},
  {Lupton}, {Schlegel}, {Finkbeiner}, {Padmanabhan}, {Bond}, {Sesar},
  {Rockosi}, {Knapp}, {Gunn}, {Sumi}, {Schneider}, {Barentine}, {Brewington},
  {Brinkmann}, {Fukugita}, {Harvanek}, {Kleinman}, {Krzesinski}, {Long},
  {Neilsen}, {Nitta}, {Snedden}, \& {York}}]{Juric08}
{Juri{\'c}}, M., {Ivezi{\'c}}, {\v Z}., {Brooks}, A., {Lupton}, R.~H.,
  {Schlegel}, D., {Finkbeiner}, D., {Padmanabhan}, N., {Bond}, N., {Sesar}, B.,
  {Rockosi}, C.~M., {Knapp}, G.~R., {Gunn}, J.~E., {Sumi}, T., {Schneider},
  D.~P., {Barentine}, J.~C., {Brewington}, H.~J., {Brinkmann}, J., {Fukugita},
  M., {Harvanek}, M., {Kleinman}, S.~J., {Krzesinski}, J., {Long}, D.,
  {Neilsen}, Jr., E.~H., {Nitta}, A., {Snedden}, S.~A., \& {York}, D.~G. 2008,
  \apj, 673, 864

\bibitem[{{Kalirai} {et~al.}(2006){Kalirai}, {Guhathakurta}, {Gilbert},
  {Reitzel}, {Majewski}, {Rich}, \& {Cooper}}]{Kalirai06}
{Kalirai}, J.~S., {Guhathakurta}, P., {Gilbert}, K.~M., {Reitzel}, D.~B.,
  {Majewski}, S.~R., {Rich}, R.~M., \& {Cooper}, M.~C. 2006, \apj, 641, 268

\bibitem[{Karachentsev {et~al.}(2000)Karachentsev, Karachentseva, Kudrya,
  Makarov, \& Parnovsky}]{Kara00}
Karachentsev, I.~D., Karachentseva, V.~E., Kudrya, Y.~N., Makarov, D.~I., \&
  Parnovsky, S.~L. 2000, Bull.~Special Astrophys.~Obs., 50, 5

\bibitem[{Karachentsev {et~al.}(1993)Karachentsev, Karachentseva, \&
  Parnovskij}]{Karachentsev93}
Karachentsev, I.~D., Karachentseva, V.~E., \& Parnovskij, S.~L. 1993,
  Astronomische Nachrichten, 314, 97

\bibitem[{{Kazantzidis} {et~al.}(2007){Kazantzidis}, {Bullock}, {Zentner},
  {Kravtsov}, \& {Moustakas}}]{Kaz07}
{Kazantzidis}, S., {Bullock}, J.~S., {Zentner}, A.~R., {Kravtsov}, A.~V., \&
  {Moustakas}, L.~A. 2007, ArXiv e-prints, 708

\bibitem[{Kelson {et~al.}(2000)Kelson, Illingworth, van Dokkum, \&
  Franx}]{Kelson00}
Kelson, D.~D., Illingworth, G.~D., van Dokkum, P.~G., \& Franx, M. 2000, \apj,
  531, 159

\bibitem[{{Koch} {et~al.}(2007){Koch}, {Rich}, {Reitzel}, {Mori}, {Loh},
  {Ibata}, {Martin}, {Chapman}, {Ostheimer}, {Majewski}, \& {Grebel}}]{Koch07}
{Koch}, A., {Rich}, R.~M., {Reitzel}, D.~B., {Mori}, M., {Loh}, Y.-S., {Ibata},
  R., {Martin}, N., {Chapman}, S.~C., {Ostheimer}, J., {Majewski}, S.~R., \&
  {Grebel}, E.~K. 2007, Astronomische Nachrichten, 328, 653

\bibitem[{Kregel \& van~der Kruit(2005)}]{Kregel05d}
Kregel, M., \& van~der Kruit, P.~C. 2005, \mnras, 358, 481

\bibitem[{Kroupa(2002)}]{Kroupa02}
Kroupa, P. 2002, \mnras, 330, 707

\bibitem[{Martin {et~al.}(2004)Martin, Ibata, Bellazzini, Irwin, Lewis, \&
  Dehnen}]{Martin04}
Martin, N.~F., Ibata, R.~A., Bellazzini, M., Irwin, M.~J., Lewis, G.~F., \&
  Dehnen, W. 2004, \mnras, 348, 12

\bibitem[{Matthews \& Wood(2001)}]{Matthews01}
Matthews, L.~D., \& Wood, K. 2001, \apj, 548, 150

\bibitem[{Mishenina {et~al.}(2004)Mishenina, Soubiran, Kovtyukh, \&
  Korotin}]{Mishenina04}
Mishenina, T.~V., Soubiran, C., Kovtyukh, V.~V., \& Korotin, S.~A. 2004, \aap,
  418, 551

\bibitem[{Morrison {et~al.}(1997)Morrison, Miller, Harding, Stinebring, \&
  Boroson}]{Morrison97}
Morrison, H.~L., Miller, E.~D., Harding, P., Stinebring, D.~R., \& Boroson,
  T.~A. 1997, \aj, 113, 2061

\bibitem[{Mould(2005)}]{Mould05}
Mould, J. 2005, \aj, 129, 698

\bibitem[{Navarro {et~al.}(2004)Navarro, Helmi, \& Freeman}]{Navarro04}
Navarro, J.~F., Helmi, A., \& Freeman, K.~C. 2004, \apjl, 601, L43

\bibitem[{Neeser {et~al.}(2002)Neeser, Sackett, De~Marchi, \&
  Paresce}]{Neeser02}
Neeser, M.~J., Sackett, P.~D., De~Marchi, G., \& Paresce, F. 2002, \aap, 383,
  472

\bibitem[{Nissen(1995)}]{Nissen95}
Nissen, P.~E. 1995, in IAU Symp. 164: Stellar Populations, 109--+

\bibitem[{Nissen {et~al.}(2003)Nissen, Chen, Asplund, \& Max}]{Nissen03}
Nissen, P.~E., Chen, Y., Asplund, M., \& Max, P. 2003, Elemental Abundances in
  Old Stars and Damped Lyman-$\alpha$ Systems, 25th meeting of the IAU, Joint
  Discussion 15, 22 July 2003, Sydney, Australia, 15

\bibitem[{Ojha(2001)}]{Ojha01}
Ojha, D.~K. 2001, \mnras, 322, 426

\bibitem[{Osterbrock {et~al.}(1997)Osterbrock, Fulbright, \& Bida}]{Oster97}
Osterbrock, D.~E., Fulbright, J.~P., \& Bida, T.~A. 1997, \pasp, 109, 614

\bibitem[{Osterbrock {et~al.}(1996)Osterbrock, Fulbright, Martel, Keane,
  Trager, \& Basri}]{Oster96}
Osterbrock, D.~E., Fulbright, J.~P., Martel, A.~R., Keane, M.~J., Trager,
  S.~C., \& Basri, G. 1996, \pasp, 108, 277

\bibitem[{Parker {et~al.}(2004)Parker, Humphreys, \& Beers}]{Parker04}
Parker, J.~E., Humphreys, R.~M., \& Beers, T.~C. 2004, \aj, 127, 1567

\bibitem[{Pohlen {et~al.}(2004)Pohlen, Balcells, L\"utticke, \&
  Dettmar}]{Pohlen04}
Pohlen, M., Balcells, M., L\"utticke, R., \& Dettmar, R.-J. 2004, \aap, 422,
  465

\bibitem[{Prochaska {et~al.}(2000)Prochaska, Naumov, Carney, McWilliam, \&
  Wolfe}]{Prochaska00}
Prochaska, J.~X., Naumov, S.~O., Carney, B.~W., McWilliam, A., \& Wolfe, A.~M.
  2000, \aj, 120, 2513

\bibitem[{Quinn {et~al.}(1993)Quinn, Hernquist, \& Fullagar}]{Quinn93}
Quinn, P.~J., Hernquist, L., \& Fullagar, D.~P. 1993, \apj, 403, 74

\bibitem[{Ram\'{\i}rez {et~al.}(2007)Ram\'{\i}rez, Allende~Prieto, \&
  Lambert}]{Ram07}
Ram\'{\i}rez, I., Allende~Prieto, C., \& Lambert, D.~L. 2007, \aap, 465, 271

\bibitem[{Reid \& Majewski(1993)}]{Reid93}
Reid, N., \& Majewski, S.~R. 1993, \apj, 409, 635

\bibitem[{Rix \& White(1992)}]{Rix92}
Rix, H., \& White, S.~D.~M. 1992, \mnras, 254, 389

\bibitem[{Robin {et~al.}(1996)Robin, Haywood, Creze, Ojha, \&
  Bienayme}]{Robin96}
Robin, A.~C., Haywood, M., Creze, M., Ojha, D.~K., \& Bienayme, O. 1996, \aap,
  305, 125

\bibitem[{Seth {et~al.}(2007)Seth, De~Jong, Dalcanton, \& the
  GHOSTS~team}]{Seth07}
Seth, A., De~Jong, R., Dalcanton, J., \& the GHOSTS~team. 2007, ArXiv
  Astrophysics e-prints

\bibitem[{Seth {et~al.}(2005)Seth, Dalcanton, \& de~Jong}]{Seth05b}
Seth, A.~C., Dalcanton, J.~J., \& de~Jong, R.~S. 2005, \aj, 130, 1574

\bibitem[{Shaw \& Gilmore(1989)}]{Shaw89}
Shaw, M.~A., \& Gilmore, G. 1989, \mnras, 237, 903

\bibitem[{Simkin(1974)}]{Simkin74}
Simkin, S.~M. 1974, \aap, 31, 129

\bibitem[{Soubiran {et~al.}(2003)Soubiran, Bienaym\'e, \& Siebert}]{Soubiran03}
Soubiran, C., Bienaym\'e, O., \& Siebert, A. 2003, \aap, 398, 141

\bibitem[{Statler(1995)}]{Statler95}
Statler, T. 1995, \aj, 109, 1371

\bibitem[{Statler(1988)}]{Statler88}
Statler, T.~S. 1988, \apj, 331, 71

\bibitem[{Tautvai{\v s}ien{\.e} {et~al.}(2001)Tautvai{\v s}ien{\.e},
  Edvardsson, Tuominen, \& Ilyin}]{Taut01}
Tautvai{\v s}ien{\.e}, G., Edvardsson, B., Tuominen, I., \& Ilyin, I. 2001,
  \aap, 380, 578

\bibitem[{Tikhonov \& Galazutdinova(2005)}]{Tikhonovo5b}
Tikhonov, N.~A., \& Galazutdinova, O.~A. 2005, Astrophysics, 48, 221

\bibitem[{Tikhonov {et~al.}(2005)Tikhonov, Galazutdinova, \&
  Drozdovsky}]{Tikhonov05}
Tikhonov, N.~A., Galazutdinova, O.~A., \& Drozdovsky, I.~O. 2005, \aap, 431,
  127

\bibitem[{Tonry \& Davis(1979)}]{Tonry79}
Tonry, J., \& Davis, M. 1979, \aj, 84, 1511

\bibitem[{Tsikoudi(1979)}]{Tsikoudi79}
Tsikoudi, V. 1979, \apj, 234, 842

\bibitem[{van~der Kruit(1984)}]{Kruit84}
van~der Kruit, P.~C. 1984, \aap, 140, 470

\bibitem[{van Dokkum {et~al.}(1994)van Dokkum, Peletier, de~Grijs, \&
  Balcells}]{vanDokkum94}
van Dokkum, P.~G., Peletier, R.~F., de~Grijs, R., \& Balcells, M. 1994, \aap,
  286, 415

\bibitem[{Velazquez \& White(1999)}]{Velazquez99}
Velazquez, H., \& White, S.~D.~M. 1999, \mnras, 304, 254

\bibitem[{Villumsen(1985)}]{Villumsen85}
Villumsen, J.~V. 1985, \apj, 290, 75

\bibitem[{Walker {et~al.}(1996)Walker, Mihos, \& Hernquist}]{Walker96}
Walker, I.~R., Mihos, J.~C., \& Hernquist, L. 1996, \apj, 460, 121

\bibitem[{Wu {et~al.}(2002)Wu, Burstein, Deng, Zhou, Shang, Zheng, Chen, Su,
  Windhorst, Chen, Zou, Xia, Jiang, Ma, Xue, Zhu, Cheng, Byun, Chen, Deng, Fan,
  Fang, Kong, Li, Lin, Lu, Sun, Tsay, Xu, Yan, Zhao, \& Zheng}]{Wu02}
Wu, H., Burstein, D., Deng, Z., Zhou, X., Shang, Z., Zheng, Z., Chen, J., Su,
  H., Windhorst, R.~A., Chen, W., Zou, Z., Xia, X., Jiang, Z., Ma, J., Xue, S.,
  Zhu, J., Cheng, F., Byun, Y., Chen, R., Deng, L., Fan, X., Fang, L., Kong,
  X., Li, Y., Lin, W., Lu, P., Sun, W., Tsay, W., Xu, W., Yan, H., Zhao, B., \&
  Zheng, Z. 2002, \aj, 123, 1364

\bibitem[{Xilouris {et~al.}(1999)Xilouris, Byun, Kylafis, Paleologou, \&
  Papamastorakis}]{Xilouris99}
Xilouris, E.~M., Byun, Y.~I., Kylafis, N.~D., Paleologou, E.~V., \&
  Papamastorakis, J. 1999, \aap, 344, 868

\bibitem[{Yoachim \& Dalcanton(2005)}]{Yoachim05}
Yoachim, P., \& Dalcanton, J.~J. 2005, \apj, 624, 701

\bibitem[{Yoachim \& Dalcanton(2006)}]{Yoachim06}
---. 2006, \aj, 131, 226

\end{thebibliography}

\end{document}